\documentclass[fleqn,usenatbib]{mnras}

\usepackage{newtxtext,newtxmath}

\usepackage[T1]{fontenc}
\usepackage{ae,aecompl}

\usepackage{xcolor}

\usepackage{graphicx}	

\usepackage{caption}
\usepackage{subcaption}
\newcommand{\e}[1]{\ensuremath{\times 10^{#1}}}

\title[Parameter estimation in sGRB plateaux]{Inferring properties of neutron stars born in short gamma-ray bursts with a plerion-like X-ray plateau}

\author[L. C. Strang and A. Melatos]{
L. C. Strang,$^{1, 2}$\thanks{E-mail: lstrang@student.unimelb.edu.au}
A. Melatos,$^{1, 2}$
N. Sarin,$^{2, 3}$
P. D. Lasky$^{2,3}$
\\
$^{1}$School of Physics, University of Melbourne, Parkville, VIC 3010 Australia\\
$^{2}$ Australian Research Council Centre of Excellence for Gravitational Wave Discovery (OzGrav)\\
$^{3}$ School of Physics and Astronomy, Monash University, Clayton, VIC 3800 Australia\\
}

\date{Accepted 2021 July 27. Received 2021 July 08; in original form 2020 December 13}

\pubyear{2021}

\begin{document}
\label{firstpage}
\pagerange{\pageref{firstpage}--\pageref{lastpage}}
\maketitle

\begin{abstract}
Time-resolved spectra of six short gamma-ray bursts (sGRBs), measured by the {\em Swift} telescope, are used to estimate the parameters of a plerion-like model of the X-ray afterglow.
The unshrouded, optically thin component of the afterglow is modelled as emanating from an expanding bubble of relativistic, shock-accelerated electrons fuelled by a central object.
The electrons are injected with a power-law distribution and cool mainly by synchrotron losses.
   We compute posteriors for model parameters describing the central engine (e.g. spin frequency at birth, magnetic field strength) and shock acceleration (e.g. power-law index, minimum injection energy).
  It is found that the central engine is compatible with a millisecond magnetar, and the shock physics is compatible with what occurs in Galactic supernova remnants, assuming standard magnetic field models for the magnetar wind.
Separately, we allow the magnetic field to vary arbitrarily and infer that it is roughly constant and lower in magnitude than the wind-borne extension of the inferred magnetar field.
This may be due to the expansion history of the bubble, or the magnetization of the circumstellar environment of the sGRB progenitor.
\end{abstract}




\section{Introduction}
\label{sec:introduction}

Among the $121$ short gamma-ray bursts (sGRBs) observed by the Neil Gehrels \textit{Swift} telescope \citep{Gehrels2004}, approximately half display long-lived emission in the X-ray band (0.3 keV--10 keV) lasting up to $\sim 10^5 \, {\rm s}$ after the burst \citep{Rowlinson2013}.
The X-ray lightcurves of sGRBs with long-lived emission often divide into three stages, termed `canonical' by various authors  \citep{Zhang2006,Nousek2006}: an initial decay, a flat plateau, and a final decay.

Neutron star mergers are confirmed as progenitors of some sGRBs \citep{Abbott2017a,Abbott2017b}, but questions remain about the origin of the X-ray plateau and the evolution of the post-sGRB remnant.
Depending on the neutron star equation of state and progenitor mass \citep{Lattimer2001}, the compact remnant may be a black hole or a neutron star.
A rapidly rotating, centrifugally supported neutron star is one possible source of the long-lived X-ray emission \citep{Dai1998,Zhang2001,fan2005late,Gompertz2013,Rowlinson2013,Lasky2017}, if its rotational energy is somehow converted to X-rays.
Such an object, if it exceeds the maximum Tolman-Oppenheimer-Volkoff mass of a stable, non-rotating neutron star \citep{Oppenheimer1939}, collapses to a black hole once it loses sufficient rotational energy \citep{Cook1994b, Cook1994a}.
The compact object is expected to be surrounded by ejecta from the collision and, potentially, additional mass outflow from the central compact object \citep{Davies1994,Rosswog1999,Li1998,Metzger2008}.
Some authors have modelled the X-ray and optical emission from a quasispherical, optically thick shroud of material surrounding the neutron star \citep{Metzger2014,Siegel2016,Yu2013}.
In this scenario, any X-rays produced within the remnant are trapped, until the shroud becomes optically thin at X-ray frequencies hours or days after the burst.
Other authors have modelled the X-ray emission by assuming that it emanates directly from the central engine \citep{Rowlinson2013,Lasky2017,sarin2020gravitational} or is produced via radiative losses from interactions with the surrounding environment \citep{2011A&A...526A.121D,2018ApJ...869..155S,sarin2020interpreting}, and does not intersect much of the shroud, e.g. because the shroud has `holes' due to a disk-jet structure \citep{Strang2019}.
Other models have sought to explain the emission by invoking internal shocks such as fireball models \citep{Piran1999}.
Both fireball models and central engine models are capable of explaining some (but not all) features of X-ray plateaux.
\citet{sarin2019x} demonstrated that GRB140903A and GRB130603B favour a simple magnetar model, making these two sGRBs particularly suited to investigating the plerion-inspired model in this paper.

Ongoing injection of energetic electrons into a magnetized bubble with adiabatic and synchrotron cooling can produce an X-ray plateau in a manner similar to plerionic supernova remnants \citep{Pacini1973,Strang2019}.
The goal of this paper is to investigate the spectral properties of the plerion model for sGRB X-ray plateaux presented by \citet{Strang2019}, and to use the spectra to estimate the spin and magnetization of the neutron star.
We apply the model to a sample of six sGRB spectra from \textit{Swift} \citep{Evans2009} using Bilby \citep{Ashton2019} as a framework for parameter estimation.
In Section \ref{sec:model}, we review the key features of the model and emphasize the respects in which it is idealized.
In Section \ref{sec:analysis}, we introduce the data and discuss the fitting procedure.
We present and discuss the results of the parameter estimation in Sections \ref{sec:results} and \ref{sec:fullspec} for point-in-time spectra and spectral evolution.
Astrophysical implications are discussed briefly in Section \ref{sec:conclusion}.
In Appendix \ref{sec:modelcf}, we discuss the analogy between the model presented here and the model in \citet{Pacini1973}.
In Appendix \ref{sec:tesc}, we consider the effects of particles escaping, e.g. through holes in the shroud.
Broadly speaking, the spectrum is modified by $\lesssim 5$, which is small compared to other systematic uncertainties in the problem.
Finally, in Appendix \ref{sec:pheldef}, we discuss the effects of photoelectric absorption below 1 keV.

\section{Plerionic emission}
\label{sec:model}

In the plerion model presented by \citet{Strang2019}, the central engine is a millisecond magnetar which injects a wind of relativistic electrons into a magnetized, expanding bubble confined by the interstellar medium, which is shock-heated by the sGRB blast wave.
The physical basis and mathematical formulation of the model closely resemble classic treatments of plerion-type young supernova remnants \citep{Pacini1973}, with the millisecond magnetar replacing an ordinary neutron star.
A more detailed comparison of the model in this paper and the model in \citet{Pacini1973} is presented in Appendix \ref{sec:modelcf}.

\subsection{Shock-accelerated electrons}
\label{sec:bubble}

The magnetar spins down by magnetic dipole braking (i.e. braking index $n=3$), if we neglect the gravitational radiation reaction torque.
The spin-down luminosity is deposited into the surrounding bubble by a relativistic magnetized wind in the form of shock-accelerated electrons with a power-law energy spectrum.
The relativistic sGRB blast wave sweeps up the shock-heated interstellar material into a thin shell, which defines the outer radius of the bubble.
The inner radius is determined by balancing the static pressure in the bubble against the ram pressure of the electrons in the magnetar wind.
We approximate the bubble of relativistic electrons as a thin shell at radius $r_b = vt$, where $r_b$ is the radius of the blast wave, $t$ is the time since the shock began, and $v$ is the expansion velocity of the shock.
The electrons lose energy by adiabatic and synchrotron cooling, as the magnetized bubble expands, with synchrotron cooling dominating at all relevant energies and time-scales for this work \citep{Strang2019}.
We calculate the evolution of the energy spectrum due to injection and cooling and hence the light curve and spectral evolution of the sGRB afterglow, including the X-ray plateau.

In the presence of optically-thick merger ejecta, some fraction $\epsilon$ of the synchrotron radiation is transmitted through holes in the ejecta.
If the shroud is unbroken, as treated by \citet{Yu2013, Metzger2014,Siegel2016}, we have $\epsilon = 0$ and no synchrotron radiation is transmitted until the ejecta become optically thin.
If the shroud is pierced by a jet, perforated by Rayleigh-Taylor instabilities or has holes for other reasons, we have $0 < \epsilon \leq 1$.
In this work, as in \citet{Strang2019}, we take $\epsilon = 1$ for simplicity, because our focus is on the plerionic emission.
In Appendix \ref{sec:tesc}, we briefly consider the possible effects of electrons escaping via the same mechanism.
We find that electrons escaping through holes in the shroud affects the spectrum by $\sim5\%$.

\subsection{Central engine parameters}
\label{sec:params}
The remnant is described by seven parameters: the mass ($M_*$) and radius ($R_*$) of the neutron star, the strength of the stellar magnetic field at the poles ($B_0$), the initial angular frequency of the star ($\Omega_0$), the maximum and minimum energies of electrons injected into the bubble ($E_{\pm 0}$), and the power-law injection index $a$.
If the magnetic field in the expanding shell is externally supplied and constant (e.g. the magnetic field in the interstellar medium), as opposed to an extension of the stellar field (which decreases with $r_b$ and hence $t$), we characterize it with an additional parameter $B$.
Throughout this work, we use the canonical neutron star mass and radius, $M_* = 1.4 M_\odot$ and $R_* = 10^4 {\rm \, m}$ \citep{Lattimer2001}.
These appear in combinations of powers of $M_*$ and $R_*$ (i.e. never singly) everywhere in the model.

The first four parameters are properties of the compact object itself.
Centrifugal break-up \citep{Cook1994a} requires $\Omega_0/2\pi \lesssim 10^3 \, {\rm Hz}$.
The angular velocity of the star decreases with time as
\begin{equation}
\Omega(t) = \Omega_0 \left(1+\frac{t}{\tau}\right)^{-1/2},
\label{eqn:spindown}
\end{equation}
where $\tau = \Omega_0 /(2\dot{\Omega}_0)$ is the magnetic dipole braking time initially.
We restrict $B_0$ to $B_0 \leq 10^{17} {\rm \, G}$, which contains the astrophysically plausible range $10^{8} {\rm \, G} \leq B_0 \leq 10^{16} {\rm \, G}$.
Combined, $B_0$ and $\Omega_0$ define the spin-down luminosity of the star as
\begin{equation}
\label{eqn:spdlum}
L_{\rm sd}(t) = L_0 \left(1+\frac{t}{\tau}\right)^{-2}
\end{equation}
for braking index $n=3$ \citep{Zhang2001}, where $L_0 = I\Omega_0^2/(2\tau)$ is the initial spin-down luminosity and $\tau = 3c^3\mu_0I/(4\pi\Omega_0^2 R_*^6 B_0^2)$ is the spin-down time scale. 
Specifying any two of $B_0$, $\Omega_0$, $L_0$, and $\tau$ is sufficient to uniquely specify the other two.
In Section \ref{sec:results}, we fit $L_0$ and $\tau$ and convert our results to posteriors on $B_0$ and $\Omega_0$.

If the stellar field extends through the shock into the bubble defined in Section \ref{sec:bubble}, one has $B(t) \propto B_0\Omega^2 r_b^{-1} \propto B_0 t^{-1}(1+t/\tau)^{-1}$ in a split-monopole wind \citep{Kennel1984,Strang2019}.
Then $B_0$ appears twice in the model: once in the synchrotron cooling expression as $B(t)^2 \propto B_0^2$, and once in $L_{\rm sd}$.
On the other hand, if the magnetic field in the bubble is externally supplied and constant, then the parameters $M_*$, $R_*$, $B_0$ and $\Omega_0$ only appear in $L_{\rm sd}(t)$ and not the synchrotron cooling expression $\propto B^2 \neq B_0^2$. 
Equations (\ref{eqn:spindown}) and (\ref{eqn:spdlum}) are unchanged, but the synchrotron luminosity and spectrum now depend on $B$ with no connection to $B_0$.
Henceforth, we call the plerion model with the split-monopole wind (and $B \propto B_0$) model A, and the model with an external magnetic field (and $B$ constant) model B.

The three parameters $E_{\pm 0}$ and $a$ describe the shock interaction between the wind from the central engine and its environment.
The maximum injection energy $E_{+0}$ is set by the balance between the magnetic-field-aligned electric potential created by the magnetar and the electromagnetic radiation reaction, e.g. due to curvature and/or synchrotron radiation in the magnetar's magnetosphere.
The injected electrons radiate predominantly near the minimum injection energy $E_{-0}$ for $a > 2$, so $E_{-0}$ must be high enough for the electron population to produce X-rays via synchrotron radiation \citep{Strang2019}.
For $E_{-0}$, we consider the range $10^{-7} \, {\rm erg} \leq E_{-0} \leq 10^{1} \, {\rm erg}$.
For $E_{+0}$, we consider the range $10^{-3} \, {\rm erg} \leq E_{+0} \leq 10^{2} \, {\rm erg}$ and require $E_{-0} < E_{+0}$.

\subsection{Spectral evolution}
\label{sec:obs}
Here we follow the derivation in \citet{Strang2019}.
In the absence of diffusive shock heating, e.g. by internal shocks in the bubble, the spatially-averaged electron energy distribution $N(E,t)$ in the bubble evolves according to  \citep{Pacini1973}

\begin{equation}
    \frac{\partial N(E,t)}{\partial t}  = \frac{\partial  }{\partial E} \left[ \left(\left.\frac{dE}{dt}\right|_{\text{ ad }}  + \left.\frac{dE}{dt}\right|_{\text{ syn }} \right) N(E,t)\right] + \dot{N}_{\text{inj}}(E,t),
\label{eqn:pdegen}
\end{equation}
where  $\dot{N}_{\text{inj}}(E,t)$ is the electron injection rate, and the powers in adiabatic and synchrotron cooling are given by
\begin{equation}
  \label{eqn:adiabatic}
  \left.\frac{dE}{dt}\right|_{\rm ad} = -\frac{E}{t},
\end{equation}
and
\begin{equation}
\left.\frac{dE}{dt}\right|_{\text{syn}} =  -\frac{4\sigma_T cE^2 B(t)^2}{24 \pi(m_ec^2)^2 },
\label{eqn:synchdef}
\end{equation}
respectively.
In (\ref{eqn:synchdef}), $\sigma_T$ is the Thomson cross-section and $B(t)$ is the magnetic field in the bubble at time $t$.

To calculate the synchrotron spectrum emitted by the plerion, we assume for simplicity that the electrons radiate at their characteristic frequency,
\begin{equation}
\nu_c = \frac{3}{2} \left(\frac{E}{m_e c^2}\right) ^2 \frac{e B(t)}{2\pi m_e c}.
\label{eqn:charfreq}
\end{equation}
This approximation introduces a smaller error than other approximations in the model.
It can be relaxed in later work if the model in its idealised form is not falsified by future observations.
The radiated flux density is proportional to
\begin{equation}
\label{eq:spec}
F_{\nu}(t) = N(E_\nu,t) \left.\frac{d E_\nu}{dt}\right|_{\rm syn} \frac{\partial E_\nu}{\partial \nu},
\end{equation}
where $\nu$ is the frequency of observation and $E_\nu$ is the energy obtained by solving equation (\ref{eqn:charfreq}) for $E$ given $\nu_c=\nu$.

For the luminosity, we integrate
\begin{equation}
  \label{eq:lumdef}
  L(t) = \int_{E_{\rm min}}^{E_{\rm max}} dE N(E,t) \left.\frac{dE}{dt}\right|_{\rm syn},
\end{equation}
where the energy band $E_{\rm min} \leq E \leq E_{\rm max}$ is defined by both the frequencies of interest to the observer and the physical system \citep{Strang2019}.
In this work, we restrict our attention to the $1 \, {\rm keV} < h\nu < 10 \, {\rm keV}$ band.

\section{Parameter estimation}
\label{sec:analysis}
\subsection{Data}
\label{sec:bilby}
All the data analysed in this paper are from the \textit{Swift} telescope and online data centre \citep{Gehrels2004,Evans2007,Evans2009}.
We consider a sample of six sGRBs of known redshift, summarized in Table \ref{tab:grblist}.
The neutral hydrogen column density $n_H$ is retrieved for each sky position from the HI4PI survey \citep{HI4PI-Collaboration2016}.
For each sGRB, we use a spectrum built from the \textit{Swift} online database for the time spans specified in Table \ref{tab:grblist}.
The choice of time span is justified in Sections \ref{sec:workedex} and \ref{sec:results}.
We pass each spectrum through \texttt{XSPEC} \citep{Arnaud1996} and filter out events flagged as having poor data quality.
We also make use of \texttt{XSPEC}'s `rebin' command, combining up to five (three) adjacent energy bins to produce a significance above $5\sigma$ ($2\sigma$) for the point-in-time spectra (evolving spectra) in Section \ref{sec:results} (~\ref{sec:fullspec}).
We assign units to the spectra using the unabsorbed counts-to-flux ratio provided by the \textit{Swift} online database in ${\rm erg \, cm}^{-2} \, {\rm s}^{-1}$.
Due to photoelectric absorption altering the spectrum below 1 keV, we restrict our fit to data in the range $1 \, {\rm keV} < \nu < 10 {\rm keV}$.
The main effects of photoelectric absorption are outlined briefly in appendix \ref{sec:pheldef}.

\begin{table*}
  \begin{tabular}{lllllll}
\hline
GRB & Redshift & $c_f$ (erg cm\(^{-2}\) ct\(^{-1}\))  & \(n_H ({\rm \, cm}^{-2})\) & $t_{\rm mean}$ (s) & Time interval  (s) & Ref\\
\hline
051221A & 0.55 & \(3.9\times 10^{-11}\)   & \(5.29\times 10^{20}\) & 7124 & 6000--10 000& \citet{Soderberg2006}\\
090510 & 0.90 & \(4.8\times 10^{-11}\)     & \(1.51\times 10^{20}\) &945 & 900--1000& \citet{Rau2009}\\
130603B & 0.36 & \(5.7\times 10^{-11}\)   & \(1.70\times 10^{20}\) & 677 & 600 -- 800& \citet{Melandri2013}\\
140903A & 0.35 & \(4.2\times 10^{-11}\)   & \(2.69\times 10^{20}\) &5335 &4000 -- 6000&  \citet{Capone2014}\\
150423A & 1.39 & \(4.8 \times 10^{-11}\) & \(1.74\times 10^{20}\) &445 & 100--1000& \citet{Malesani2015} \\
190627A & 1.94 & \(3.6 \times 10^{-11}\) & \(9.67\times 10^{20}\) &4693 & 4100 -- 5100 & \citet{Japelj2019} \\
\hline
\end{tabular}
\caption{Sample of sGRBs analysed in this paper. The references pertain to the redshift identification. Here $c_f$ is the ``counts to flux'' ratio and $t_{\rm mean}$ is the mean photon arrival time, both as reported by the \textit{Swift} online data centre.}
\label{tab:grblist}
\end{table*}

\subsection{Bayesian inference}
\label{sec:bayes}

We perform Bayesian inference using the python package Bilby \citep{Ashton2019}.
We choose priors uniform in $a$ and in $\log_{10}B$, $\log_{10} L_0$, $\log_{10} \tau$, and $\log_{10} E_{\pm 0}$.
It is more efficient to sample $L_0 = I\Omega_0^2/(2\tau)$ and $\tau = 3c^3\mu_0I/(4\pi\Omega_0^2 R_*^6 B_0^2)$ than the underlying $B_0$ and $\Omega_0$; we transform back into $B_0$ and $\Omega_0$ when analysing the results.
Because $L_0$ and $\tau$ are not linear functions of $B_0$ and $\Omega_0$, the implicit priors on $B_0$ and $\Omega_0$ are neither uniform nor log uniform but are instead uniform in $\log_{10}B_0^2\Omega_0^4$ and $\log_{10}B_0^{-2}\Omega_0^{-2}$.
In practice, these priors are fairly flat in the region of interest [ $10^{12} \lesssim B_0/\left(1 \, {\rm G}\right) \lesssim 10^{17}$ and $15 \lesssim \Omega_0/2\pi/\left(1 \, {\rm Hz}\right) \lesssim  10^3$ ] and taper off at lower values. 
In addition to the priors in Table \ref{tab:priors}, we apply upper bounds of $B_0 < 10^{17} \, {\rm G}$ and $\Omega_0/2\pi < 10^3 \, {\rm Hz}$.
We use the nested sampler \textsc{pymultinest} \citep{feroz2009multinest,2014A&A...564A.125B} with a Gaussian likelihood.
Specifically, for a flux measurement $\nu_i$ at time $t_i$, we have the likelihood
\begin{equation}
  \label{eqn:likelihood}
P(F_{\nu_i}, t_i| x, X) = \frac{1}{\sqrt{2\pi\sigma^2}}
                      \mathrm{exp}\left\{ \frac{-[F_{\nu_i} - F_\nu(t_i, x)]^2}{2\sigma^2}\right\},
\end{equation}
where $x$ represent the set of parameters appropriate for the model $X$ (A or B) specified in Section \ref{sec:model} and $\sigma$ is estimated from the data.

The prior ranges are summarized in Table \ref{tab:priors}.
We perform parameter estimation for two models: plerionic emission with a dipole magnetic field (model A) and plerionic emission with a constant magnetic field (model B).

\subsection{Temporal averaging}
\label{sec:workedex}

In this paper we study point-in-time spectra $F_\nu(t)$, calculated from equation (\ref{eq:spec}), and bolometric light curves $L(t)$, calculated from equation (\ref{eq:lumdef}).
In practice both quantities, especially $F_\nu(t)$, need to be averaged over time when comparing with observations; the X-ray flux of a typical sGRB afterglow is too low for a truly instantaneous spectrum to be measured.
On the other hand, the afterglow emission evolves rapidly, on time-scales as short as $L/\left|\dot{L} \right| \sim 10^2 \, {\rm s}$, while the averaging time-scale required to produce a reliable spectrum is normally longer, e.g. $T_{\rm av} \sim $ a few times $10^2 \, {\rm s}$.
One must therefore ask: is it fair to regard spectra measured with $T_{\rm av}\sim 10^2 \, {\rm s}$ (early afterglow) or $T_{\rm av} \sim 10^3$ (late afterglow) as being `instantaneous' to an acceptable approximation?

\begin{figure}
  \begin{subfigure}{0.45\textwidth}
  \includegraphics[width=\linewidth]{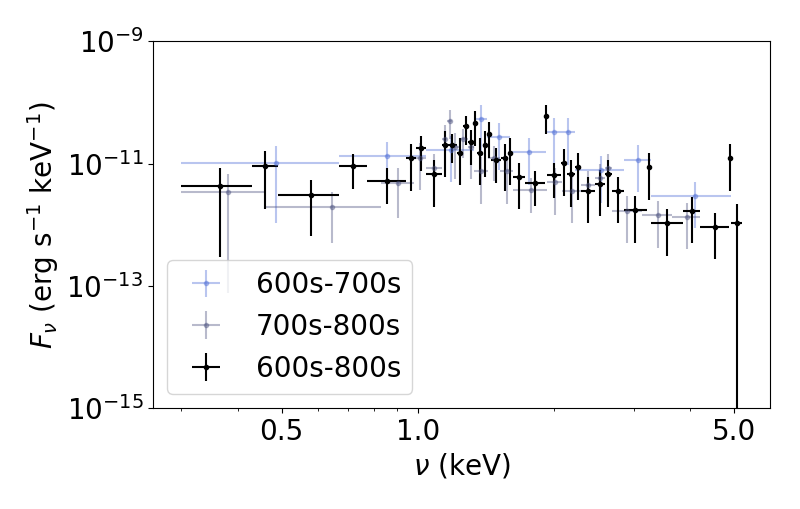}
  \caption{\label{fig:slice1}}
  \end{subfigure}
  \begin{subfigure}{0.45\textwidth}
  \includegraphics[width=\linewidth]{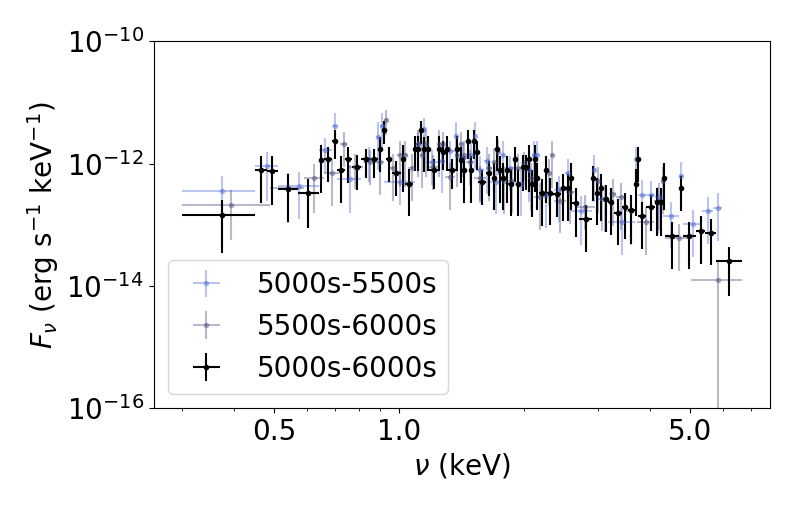}
  \caption{\label{fig:slice2}}
  \end{subfigure}
  \caption{\label{fig:sliced} Synchrotron spectral flux density $F_\nu$ (erg s$^{-1}$ keV$^{-1}$) versus frequency $\nu$ (keV). Top panel: the data points show the early-stage time-averaged spectral flux density for  $600 \leq t/(1 \, {\rm s}) \leq 700$ (blue), $700 \leq t/(1 \, {\rm s}) \leq 800$ (grey), and $600 \leq t/(1 \, {\rm s}) \leq 800$ (black). Bottom panel: late stage; $5000 \leq t/(1 {\rm s}) \leq 5500$ (blue), $5500 \leq t/(1 {\rm s}) \leq 6000$ (grey), and $5000 \leq t/(1 {\rm s}) \leq 6000$ (black).}
\end{figure}

Let us look first at the data.
Figure \ref{fig:sliced} shows six spectra measured for GRB130603B: three at early times, with $600 \leq t/(1 {\rm s}) \leq 700$, $700 \leq t/(1 {\rm s}) \leq 800$, and $600 \leq t/(1 {\rm s}) \leq 800$ (top panel), and three at late times, with $5000 \leq t/(1 {\rm s}) \leq 5500$, $5500 \leq t/(1 {\rm s}) \leq 6000$, and $5000 \leq t/(1 {\rm s}) \leq 6000$ (bottom panel).
At early times, the two half-snapshots $600 \leq t/(1 {\rm s}) \leq 700$ and $700 \leq t/(1 {\rm s}) \leq 800$ at the beginning and end of the interval agree well with the snapshot  averaged over the whole interval [ $600 \leq t/(1 {\rm s}) \leq 800$; $T_{\rm av} = 200 {\rm s}$ ].
In other words, although the source evolves throughout the interval, the shape of its spectrum does not change much, and the averaging procedure does not distort the results.
The same is true at later times, in the bottom panel of Figure \ref{fig:sliced}, when the averaging time-scale is longer ($T_{\rm av} = 10^3 {\rm s}$), but the evolution of the source is slower.

\begin{figure}
  \centering 
  \includegraphics[width=\linewidth]{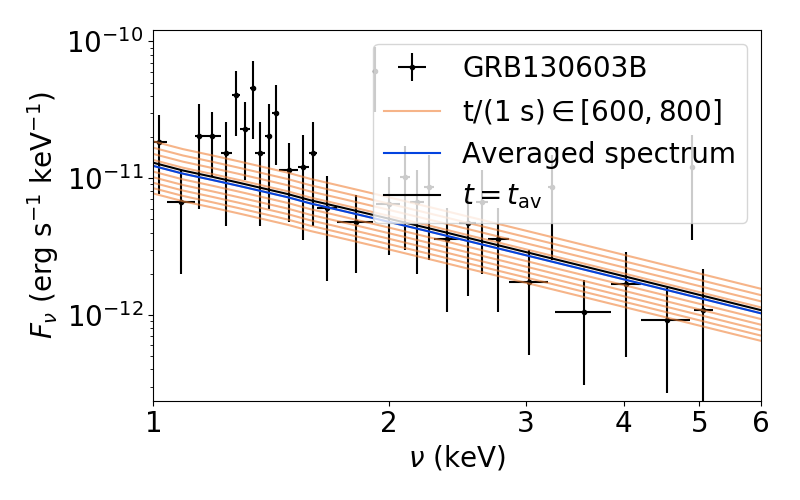}
  \caption{\label{fig:timeave} Synchrotron spectral flux $F_\nu(t)$ (erg s$^{-1}$ keV$^{-1}$) versus frequency $\nu$ (keV). Black crosses are data observed by \textit{Swift} averaged over $600 \leq t/(1 \, {\rm s}) \leq 800$. The black curve is the theoretical spectrum generated at $t = t_{\rm mean}$ for $600 \leq t/(1 \, {\rm s}) \leq 800$ using averaged random samples from the posterior. Orange curves are theoretical spectra from the same random sample as the black curve for ten equal-length intervals across $600 \leq t/(1 \, {\rm s}) \leq 800$; the blue curve is the average of the latter ten spectra. The orange curve with the brightest flux corresponds to the curve at $t = 600 \, {\rm s}$; as time progresses, the flux decreases.}
\end{figure}

Now let us look at the theory.
Equations (\ref{eqn:pdegen})--(\ref{eq:lumdef}) can be solved in closed form for $N(E, t)$ and hence $F_\nu(t)$, as demonstrated by \citet{Strang2019}.
In principle, therefore, it is possible to average $F_\nu(t)$ over the time interval in question, viz. $T_{\rm av}^{-1}\int_t^{t+T_{\rm av}} {\rm d}t' F_\nu(t')$, and compare it directly with the data as discussed above.
In practice, the integral involved in the temporal average must be done numerically, and the computation time is prohibitive for nested sampling.
We therefore verify that a theoretical point-in-time spectrum calculated at the instant $t' = t_{\rm mean}$ (where $t_{\rm mean}$, the average photon arrival time, is reported by \textit{Swift}) gives a fair approximation to the evolving spectrum throughout the interval as well as its temporal average.
Figure \ref{fig:timeave} displays the results.
Data from GRB130603B for the early-stage interval $600 \leq t/(1\,{\rm s}) \leq 800$ are passed through Bilby to generate a posterior for model A at $t_{\rm mean} = 677 \, {\rm s}$.
We then take 100 samples of the posterior and calculate the spectrum from (\ref{eqn:synchdef}) at ten equally separated instants, i.e. at $t/(1 \, {\rm s}) = 600, \, 622, \,644, \,... \,800$, and find the mean spectrum for each instant.
All ten spectra have similar shapes and lie near one another, implying that $F_\nu(t_{\rm mean})$ is a reasonable (and computationally efficient) approximation to the time-averaged theoretical spectrum.
A similar conclusion is reached for the late stage interval $5000 \leq t/(1 {\rm \, s}) \leq 6000$ (not shown).

The above results are consistent with the claim in the literature, that the spectra of sGRB afterglows do not change much at late times \citep{Evans2009}.
In the plerion picture, this occurs because the synchrotron loss time is short ($\lesssim 1 \, {\rm s}$), so the shape of $F_\nu(t)$ is dominated by the injected spectrum $\dot{N}_{\rm inj} (E, t) \propto E^{-a}$, whose energy dependence $E^{-a}$ is constant.
However, there is no fundamental reason why $\dot{N}_{\rm inj}(E, t)$ cannot change its form under some circumstances.
If observational evidence emerges that the shape of $F_\nu(t)$ evolves rapidly (faster than $T_{\rm av}$) in some sGRBs, the temporal averaging procedure described in this section does not work well in those objects.

\section{Point-in-time X-ray spectrum}
\label{sec:results}

In this section, we apply the plerion model to the spectra of the six sGRBs listed in Table \ref{tab:grblist}.
As discussed in Section \ref{sec:workedex}, we analyse point-in-time spectra constructed by averaging over a relatively narrow window bracketing the mean photon arrival time $t_{\rm mean}$, where $t_{\rm mean}$ is chosen based on the availability of detailed spectral data.
Where possible, we choose $t_{\rm mean}$ to be close to the plateau break.

\begin{center}
\begin{table}
\begin{tabular}{llll}
\hline
Parameter      & Model             & Lower bound & Upper bound\\
\hline
\(L_0\) (erg s$^{-1}$)               & B  & \(10^{37}\) & \(10^{52}\)\\
\(\tau\) (s)     & A and B  & 1 & $10^{10}$ \\
\(E_{-0}\) (erg)  &  A and B &\(10^{-7}\) & \(10^{1}\)\\
\(E_{+0}\) (erg)  & A and B &\(10^{-3}\) & \(10^{2}\)\\
\(a\)                       & A and B & 1 & 8\\
$B$ (G) &  B       &$10^{-2}$ & $10^{10}$ \\
\hline
\end{tabular}
\caption{Upper and lower bounds on priors for model A (plerion with a split monopole wind) and model B (plerion with a constant magnetic field). The priors are uniform in the logarithm of the parameter, except for $a$, whose prior is uniform. }
\label{tab:priors}
\end{table}
\end{center}

\subsection{Millisecond magnetar hypothesis}
\label{sec:params}

The plerion model studied in this work assumes that the sGRB remnant is a neutron star.
With that assumption, parameter estimation favours a millisecond magnetar over an ordinary neutron star with a weaker magnetic field and longer spin period.

\begin{figure*}
  \begin{subfigure}{0.43\textwidth}
  \includegraphics[width=\linewidth]{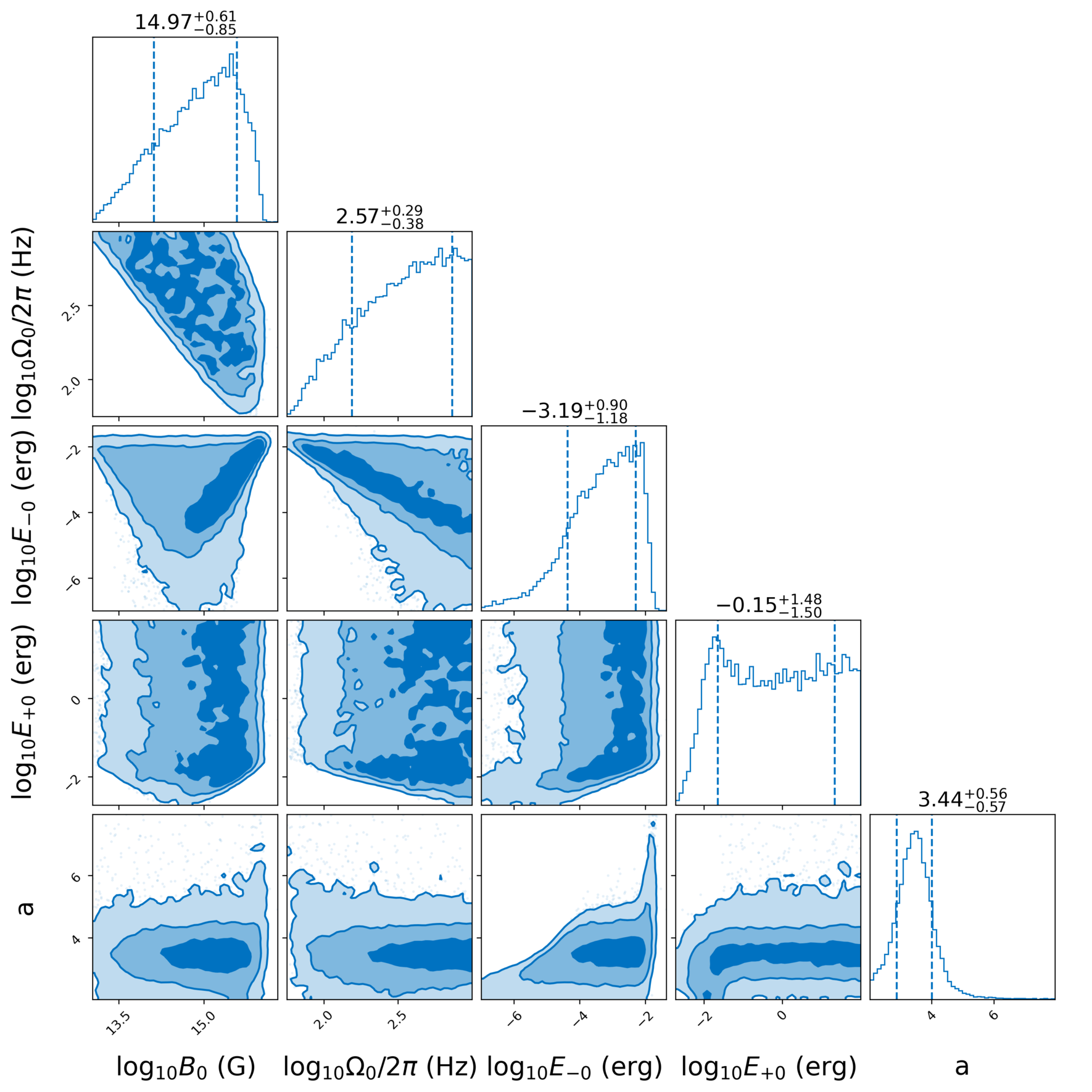}
  \caption{\label{fig:051221A_corner} GRB051221A}
  \end{subfigure}
  \begin{subfigure}{0.43\textwidth}
  \includegraphics[width=\linewidth]{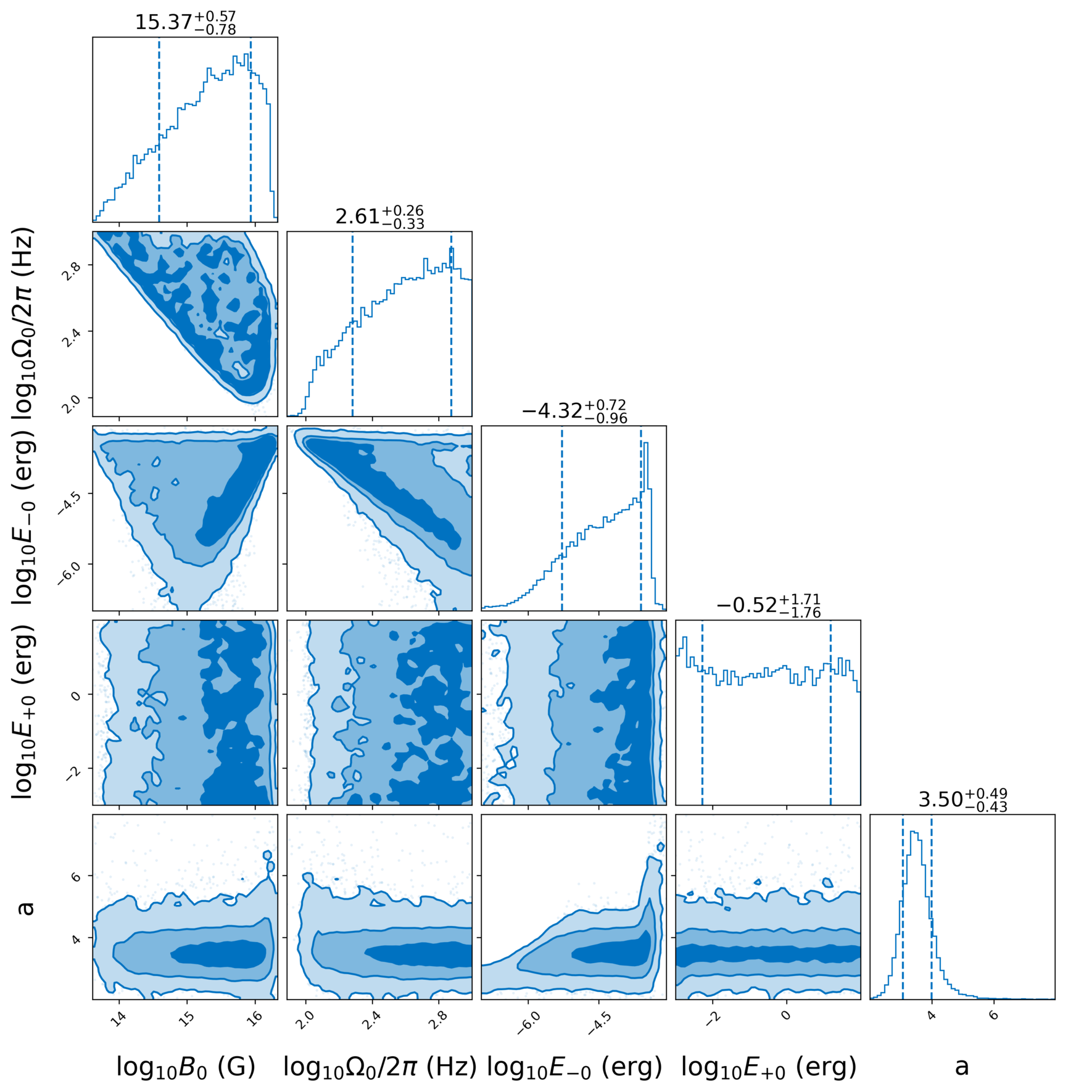}
  \caption{\label{fig:090510_corner} GRB090510}
  \end{subfigure}
  \begin{subfigure}{0.43\textwidth}
  \includegraphics[width=\linewidth]{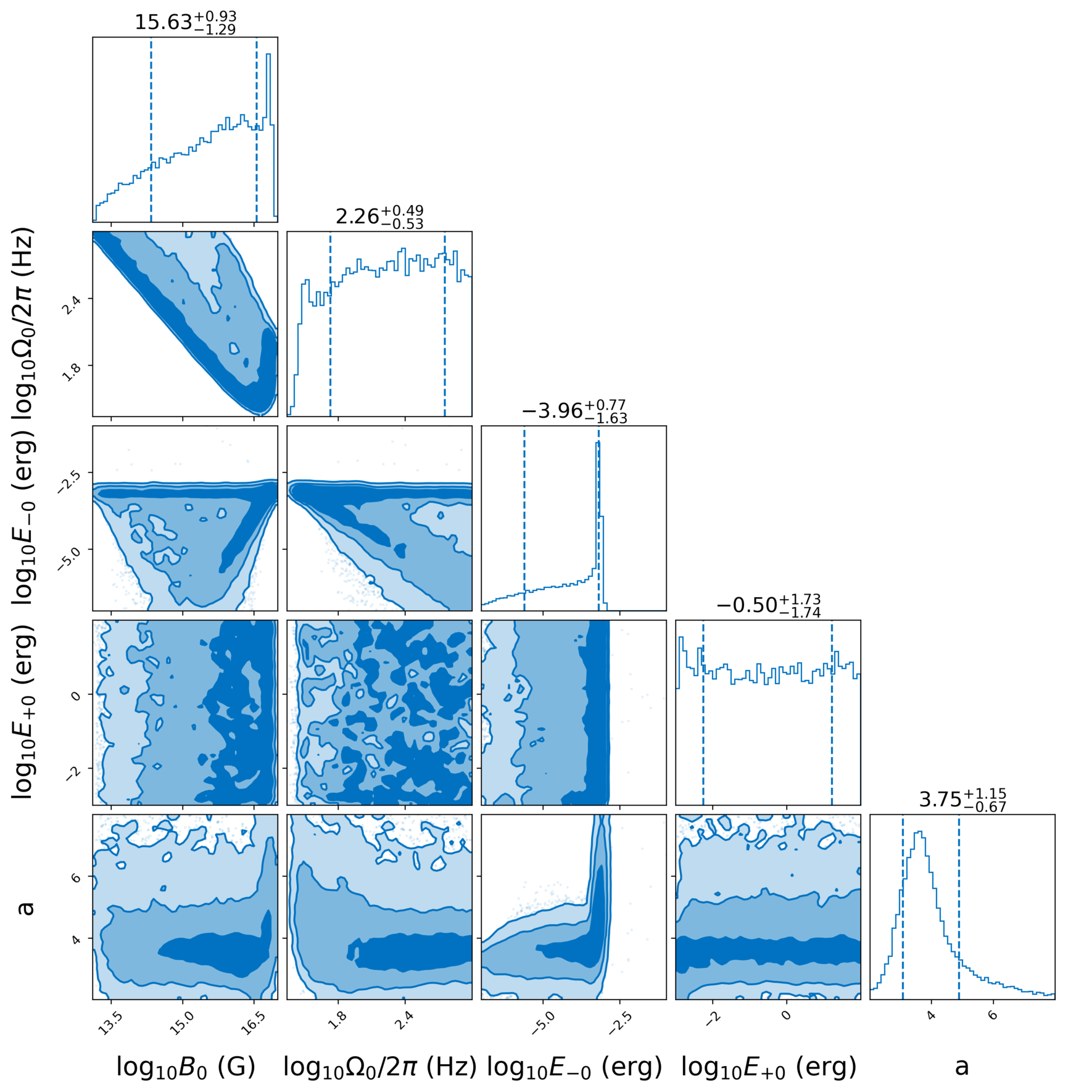}
  \caption{\label{fig:130603B_corner} GRB130603B}
  \end{subfigure}
  \begin{subfigure}{0.43\textwidth}
  \includegraphics[width=\linewidth]{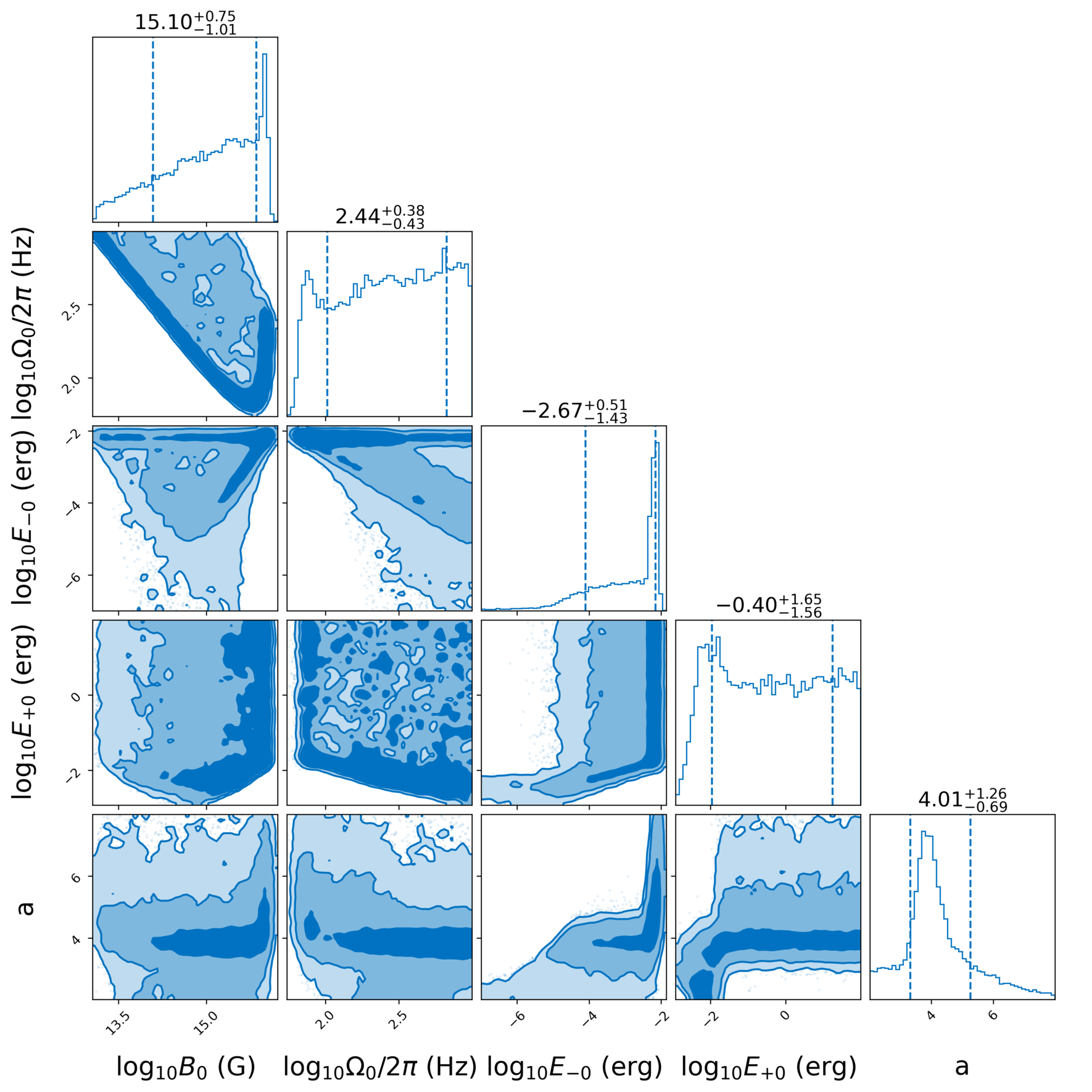}
  \caption{\label{fig:140903A_corner} GRB140903A}
  \end{subfigure}
  \begin{subfigure}{0.43\textwidth}
  \includegraphics[width=\linewidth]{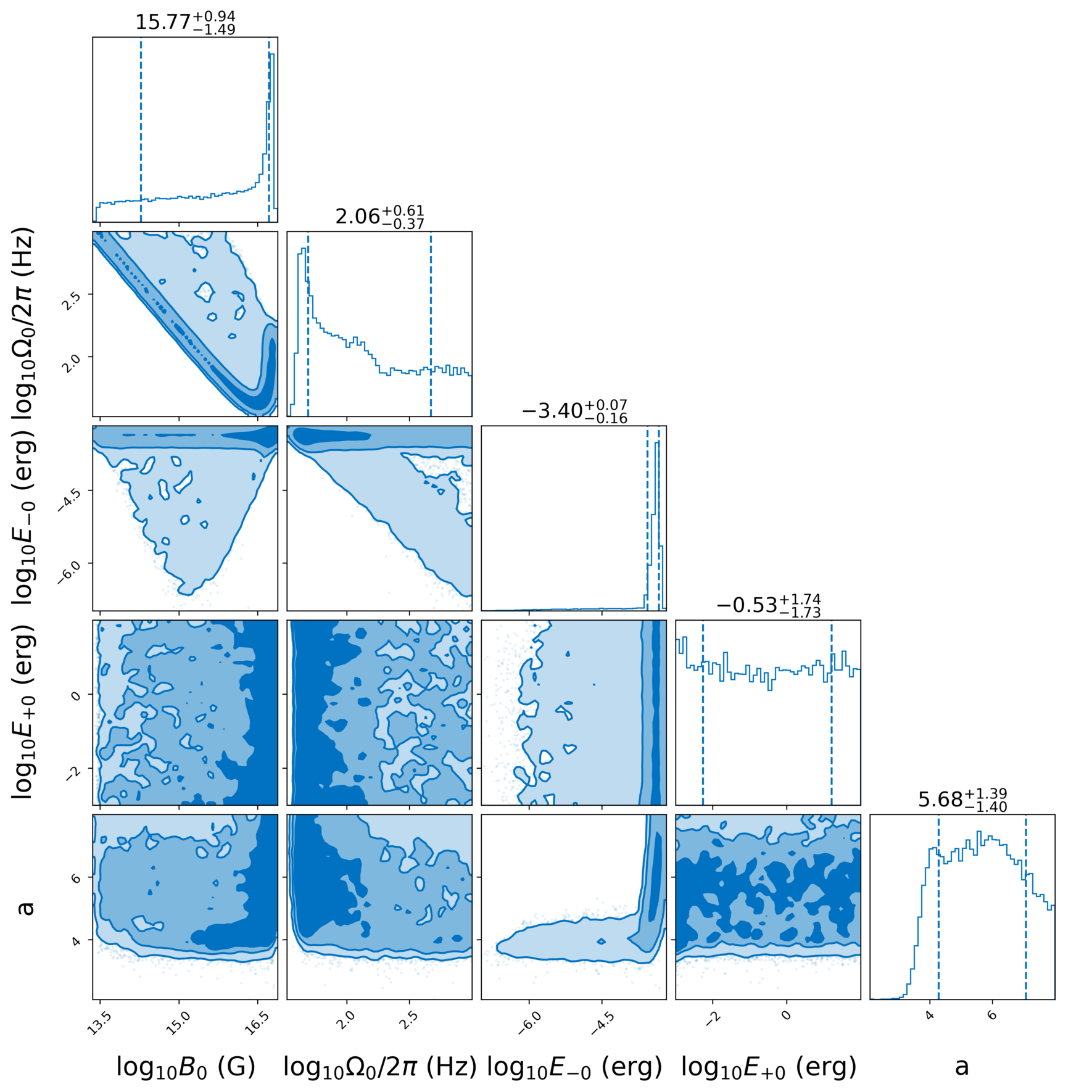}
  \caption{\label{fig:150423A_corner} GRB150423A}
  \end{subfigure}
  \begin{subfigure}{0.43\textwidth}
  \includegraphics[width=\linewidth]{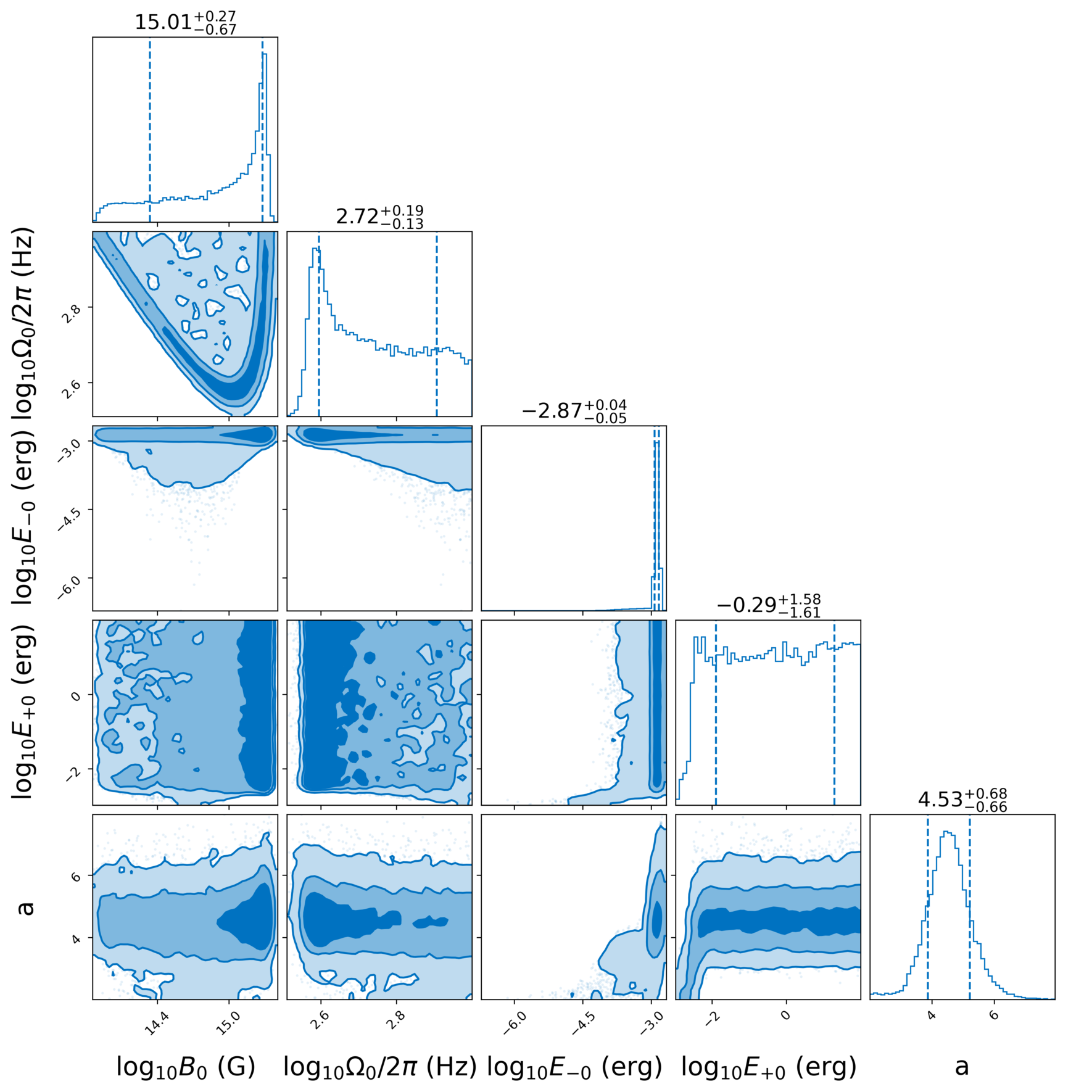}
  \caption{\label{fig:190627A_corner} GRB190627A}
  \end{subfigure}
  \caption{\label{fig:corners} Corner plots showing the posterior distributions of the plerion model A parameters $\log_{10}B_0$ (G), $\log_{10}\Omega_0$ (Hz/2$\pi$), $\log_{10}(E_{\pm 0}) \, {\rm \, (erg)}$, and $a$. Panels correspond to the six objects in Table \ref{tab:grblist}. A subset of the prior range is displayed to aid readability. }
\end{figure*}

\begin{figure*}
  \begin{subfigure}{0.43\textwidth}
  \includegraphics[width=\linewidth]{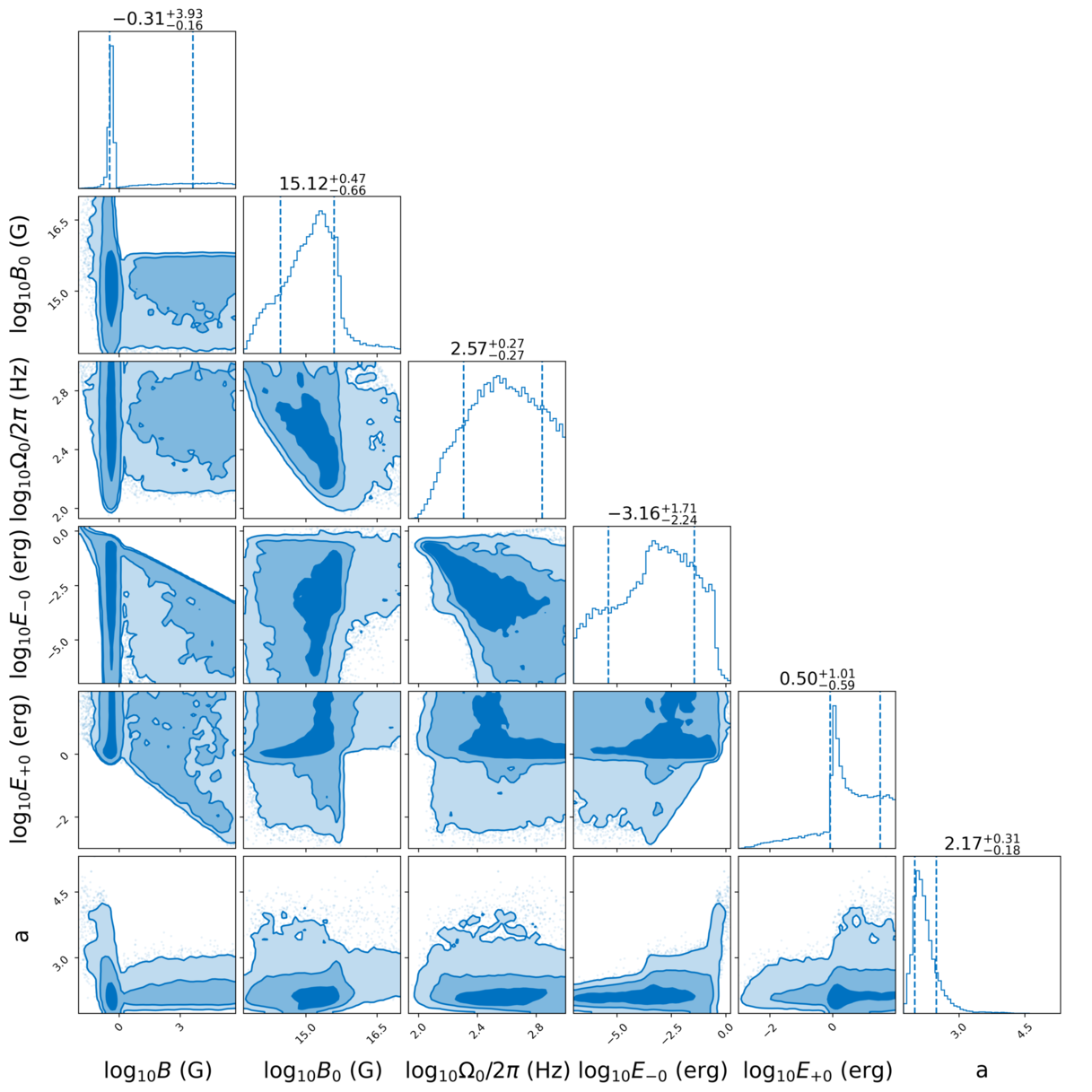}
  \caption{\label{fig:const_051221A_corner} GRB051221A}
  \end{subfigure}
  \begin{subfigure}{0.43\textwidth}
  \includegraphics[width=\linewidth]{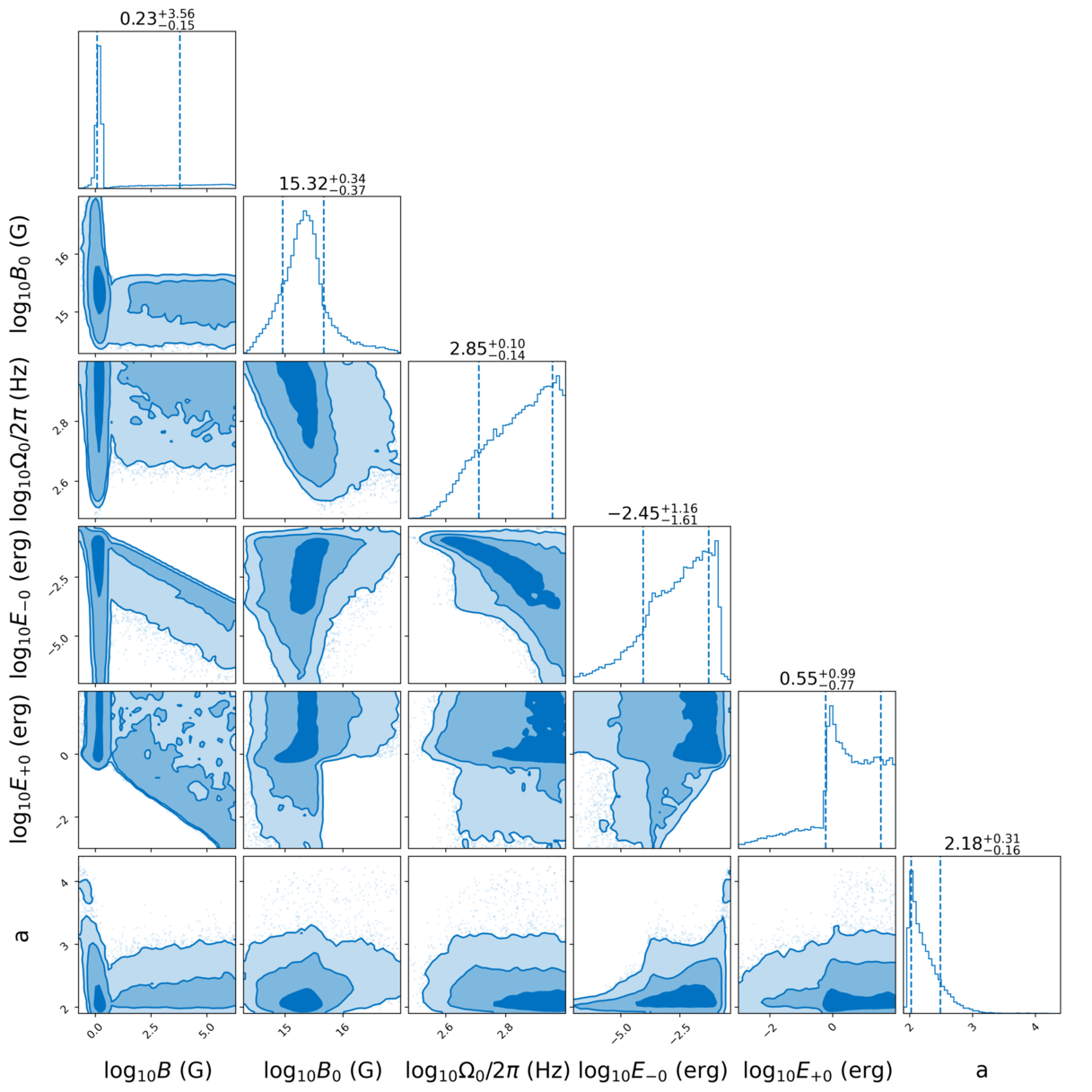}
  \caption{\label{fig:const_090510_corner} GRB090510}
  \end{subfigure}
  \begin{subfigure}{0.43\textwidth}
  \includegraphics[width=\linewidth]{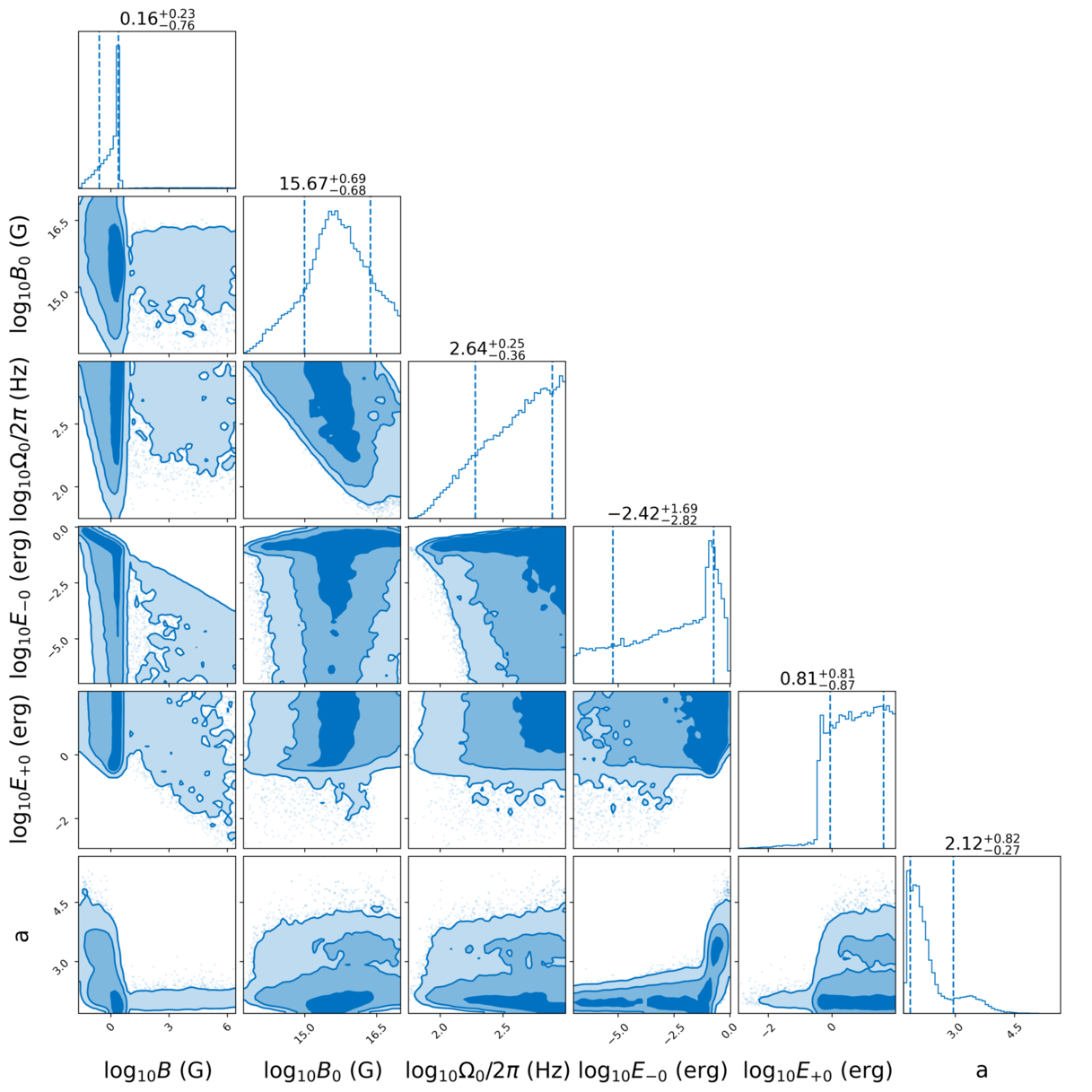}
  \caption{\label{fig:const_130603B_corner} GRB130603B}
  \end{subfigure}
  \begin{subfigure}{0.43\textwidth}
  \includegraphics[width=\linewidth]{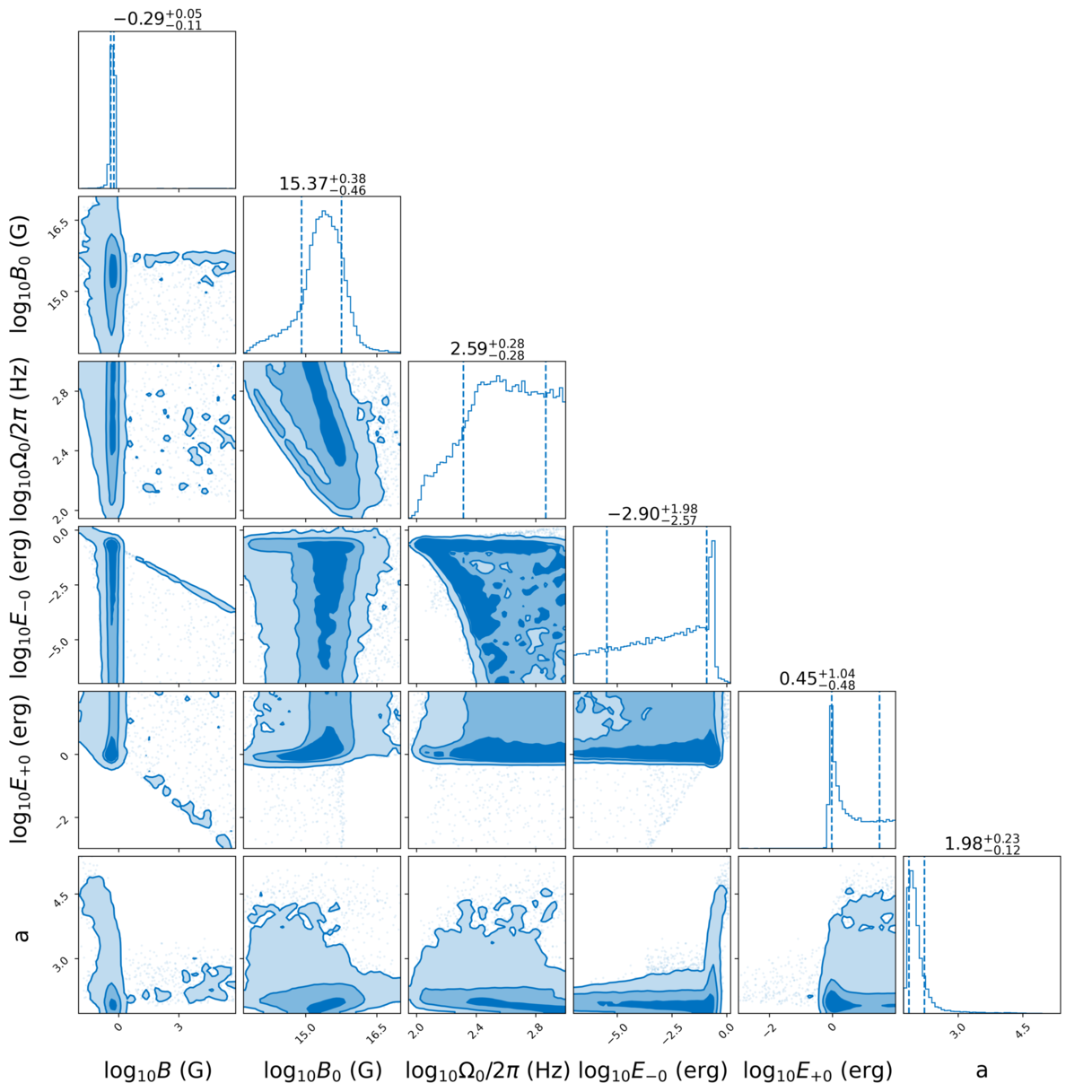}
  \caption{\label{fig:const_140903A_corner} GRB140903A}
  \end{subfigure}
  \begin{subfigure}{0.43\textwidth}
  \includegraphics[width=\linewidth]{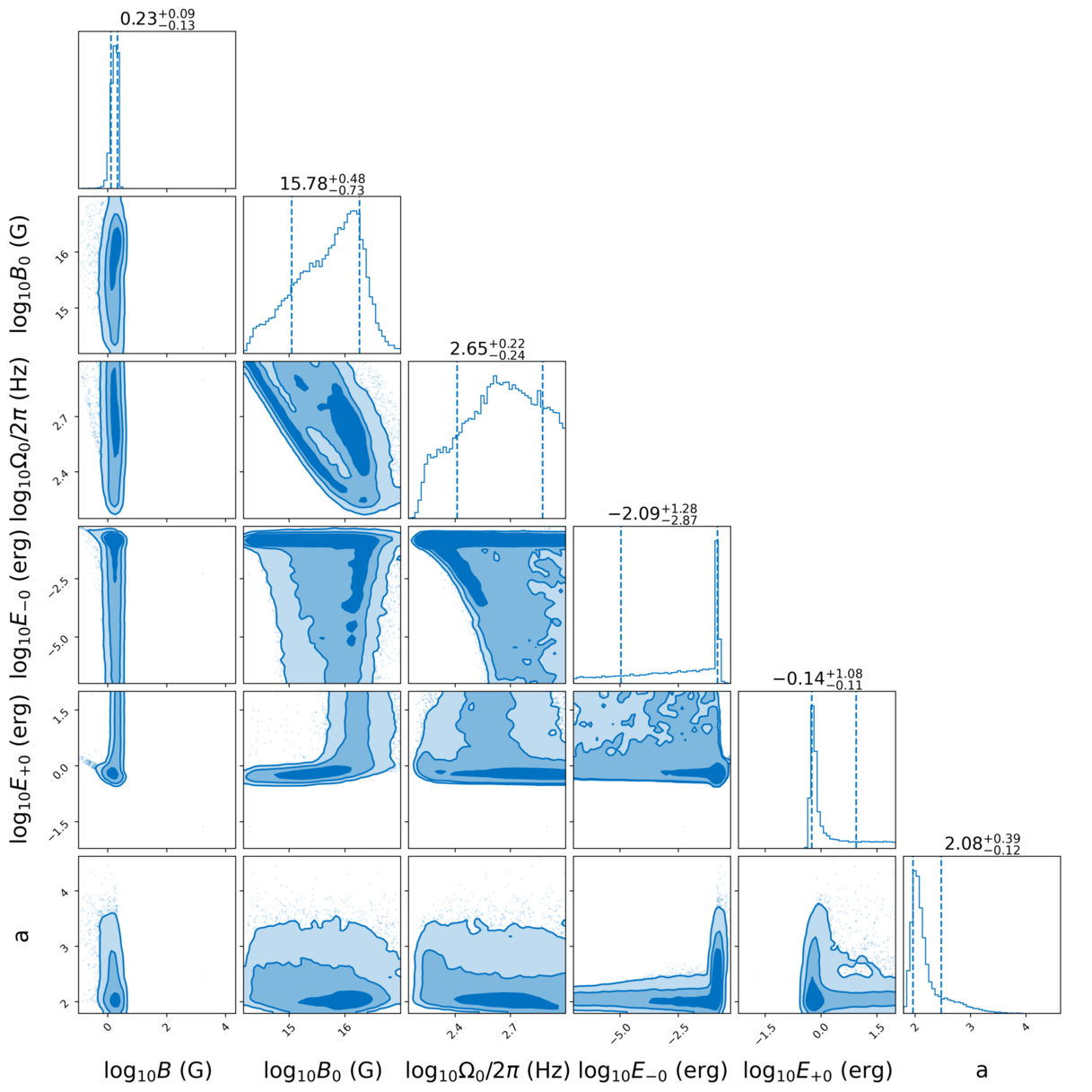}
  \caption{\label{fig:const_150423A_corner} GRB150423A}
  \end{subfigure}
  \begin{subfigure}{0.43\textwidth}
  \includegraphics[width=\linewidth]{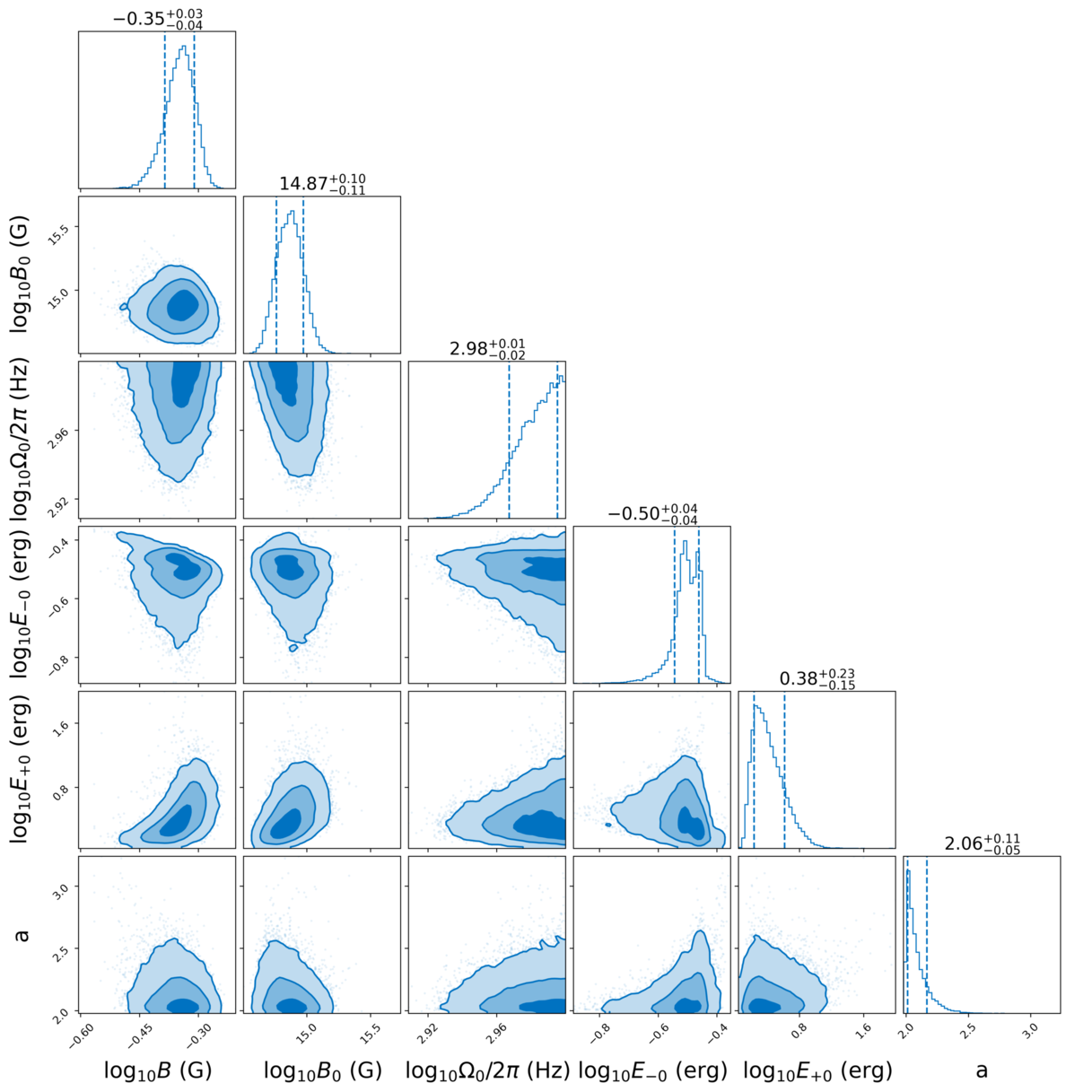}
  \caption{\label{fig:const_190627A_corner} GRB190627A}
  \end{subfigure}
  \caption{\label{fig:const_corners} Corner plots showing the posterior distributions for plerion model B. Panels correspond to six of the six objects in Table \ref{tab:grblist}.  The parameters are: $\log_{10} B$, $\log_{10}B_0$,$\log_{10}\Omega_0/2\pi$, $\log_{10}(E_{\pm 0})$, and $a$. Other model parameters are held constant. A subset of the prior domain is displayed to aid readability.}
\end{figure*}

Figures \ref{fig:corners} and \ref{fig:const_corners} display corner plots summarizing the parameter estimates for models A and B respectively for the six objects in Table \ref{tab:grblist}.
Consider model A first.
Each corner plot displays the distributions of $\log_{10}B_0$, $\log_{10}\Omega_0/2\pi$, $\log_{10}E_{-0}$, $\log_{10}E_{+0}$ and $a$, plotted on a linear axis.
The mean value and 68\% confidence interval are displayed above each marginalized posterior distribution.
The posteriors for $\log_{10}B_0$ cover the range $3\times 10^{13} \lesssim B_0/(1 \, {\rm G}) \lesssim 3 \times 10^{16}$, peak at $B_0 \gtrsim 10^{15} \, {\rm G}$, and have a mean of $B_0  \gtrsim 8 \times 10^{14} \, {\rm G}$, which is in line with a magnetar field.
The posteriors for $\log_{10}\Omega_0/2\pi$ cover the range $300 \lesssim \Omega_0/2\pi/\left(1 \, {\rm Hz}\right) \lesssim 10^3$, railing up against the physical upper bound; however, as this is approximately the centrifugal break-up frequency of a neutron star, we choose not to repeat the analysis with a larger prior range.
There are two shapes that appear in the $\log_{10}\Omega_0/2\pi$ posteriors.
The posteriors for GRB051221A and GRB090510 increase slowly from $\Omega_0/2\pi \approx 10^2 \, {\rm Hz}$ and peak at the upper bound.
In contrast, GRB130603B, GRB140903A, GRB150523A, and GRB190629A peak below $\Omega_0/2\pi \approx 10^2 \, {\rm Hz}$ and flatten for $\Omega_0/2\pi > 100 \, {\rm Hz}$.

The model A posteriors in Figure \ref{fig:corners} are skewed for $\log_{10}B_0$ for all six sGRBs, as measured by the normalized third central moment of the distribution tabulated in Table \ref{tab:skew}.
Both $\log_{10}B_0$ and $\log_{10}\Omega_0$ are correlated in the posteriors.
In the contour plots in Figure \ref{fig:corners}, the correlation appears as either a banana (for example, GRB051221A ) or a U-shape (GRB090510 and GRB190627A) in the $\log_{10}B_0$--$\log_{10}\Omega_0/2\pi$ plane.
This is not surprising; they appear together in the product $B_0^n \Omega_0^m$ ($n$, $m$ integers) in the theory, which accounts for the banana-shaped correlations.
The origin of the right arm of the U-shaped correlations is unclear

For model B, the corner plots in Figure \ref{fig:const_corners} display $\log_{10}B$, $\log_{10}B_0$, $\log_{10}\Omega_0/2\pi$, $\log_{10}E_{-0}$, $\log_{10}E_{+0}$ and $a$.
The mean value of the posterior is displayed above each marginalized distribution.
There are several trends evident across the six GRBs.
As with model A, the posteriors on $B_0$ are consistent with a millisecond magnetar, with $10^{14} \lesssim B_0/(1 \, {\rm G}) \lesssim 10^{16}$.
The parameters $B_0$ and $\Omega_0$ are correlated, with smaller $\Omega_0$ corresponding to larger $B_0$, producing a banana (GRB051221A, GRB090510, GRB130603B, and GRB190627A) or diagonal U-shape (GRB140903A and GRB150423A) in the posteriors.
The posteriors for $\log_{10}\Omega_0/2\pi$ cover the range $10 \lesssim \Omega_0/2\pi/\left(1 \, {\rm Hz}\right) \lesssim  10^3$, again railing up against the physical upper bound. 

The posteriors on $B$ cover the range $10^{-1} \lesssim B/(1 \, {\rm G}) \lesssim 1$. 
This is much stronger than the magnetic field in the interstellar medium ($\sim 10^{-6} \, {\rm G}$) but smaller than the expected field advected outwards from the central object by the relativistic outflow (i.e. the magnetic field in model A).
The posteriors for GRB051221A and GRB090510 feature a low extended plateau between $10^{0} \lesssim B/(1 \, {\rm G}) \lesssim 10^4$ which shows a correlation with the posteriors on $E_{-0}$; the same correlation is observed in GRB190627A.
The posteriors on $B$ are strongly correlated with those for $E_{-0}$ because the characteristic frequency of synchrotron radiation scales as $\nu_c \propto B E^2$.

\begin{center}

\begin{table}
\begin{tabular}{lrr}
\hline
GRB & $\log_{10}B_0$ skewness \\
\hline
051221A & -0.44 \\
090510 & -0.46 \\
130603B & -0.42 \\
140903A & -0.48  \\
150423A & -0.45 \\
190627A & -0.72 \\
\hline
\end{tabular}
\caption{\label{tab:skew} The skewness of the posteriors for $\log_{10}B_0$ in model A (split monopole wind), as measured by the third central moment divided by the cube of the variance.}
\end{table}
\end{center}

\subsection{Shock properties}
The inferred values of the particle injection parameters are also consistent in broad terms with the millisecond magnetar hypothesis.
Figures \ref{fig:corners} reveal $E_{-0}$ peaks around $E_{-0} = 10^{-3} \, {\rm erg}$ for model A and around $E_{-0} = 3\times 10^{-2} \, {\rm erg}$ for model B.
This is consistent with electrons being injected into a relativistic pulsar wind (and hence into the wind termination shock) with a radiation-reaction-limited Lorentz factor $\lesssim 3\times 10^9$ following electrostatic acceleration across a homopolar polar-cap potential $\approx 10^{21} (B_0/1\times 10^{15} {\rm \, G}) [\Omega/ (10^3 {\rm \, Hz})]^2 {\rm \, V}$ in the magnetar's magnetosphere \citep{Goldreich1969,1975ApJ...196...51R}.
For both models A and B, the posteriors for $E_{+0}$ rail against the upper bound of the prior, however, we do not increase the prior range because the upper bound is set by the radiation-reaction limit.

For model A, $E_{+0}$ returns a posterior which is uniform in the logarithm, indicating that the upper bound on the electron energy distribution does not affect the spectrum, provided it is above the maximum energy recorded.
The lower cut-off for the uniform distribution is visible in GRB051221A, GRB140903A, and GRB190627A.
For model A, $E_{-0}$ covers the range $10^{-6} \lesssim E_{-0}/(1 \, {\rm erg}) \lesssim 1$ and displays a sharp peak near $E_{-0} \approx  10^{-2} \, {\rm erg}$.
The location of the peak varies across each GRB in the range $10^{-3} \lesssim E_{-0}/({\rm 1 \, erg}) \lesssim 10^{-2}$.
For both models, $E_{+0}$ returns a posterior railing against the upper bound enforced by radiation-reaction limit.
For model B, the posterior for $E_{-0}$ covers the range $10^{-7} \lesssim E_{-0}/(1 \, {\rm erg}) \lesssim 1$ and displays a sharp peak near $E_{-0} \approx  10^{-2} \, {\rm erg}$.
In a magnetic field of $B = 1 \, {\rm G}$, this corresponds to a characteristic synchrotron frequency of $\nu_c \lesssim 0.1 \, {\rm keV}$, just below the minimum frequency observed by {\em Swift}.

Finally, we consider the injection index.
For model A, it covers the range $2 \lesssim  a \lesssim 6$, except for GRB150423A, which covers the range $4 \lesssim a \lesssim 8$.
For model B, it covers the range $1 \lesssim a \lesssim 5$.
These values are similar to those observed in Galactic supernova remnants \citep{doi:10.1146/annurev.astro.44.051905.092528}, and also consistent with the fireball model for sGRBs \citep[e.g.][]{zhang2007gamma}.

\subsection{Spectral shape}
\label{sec:spectra}

\begin{figure*}
  \begin{subfigure}{0.45\textwidth}
  \includegraphics[width=\linewidth]{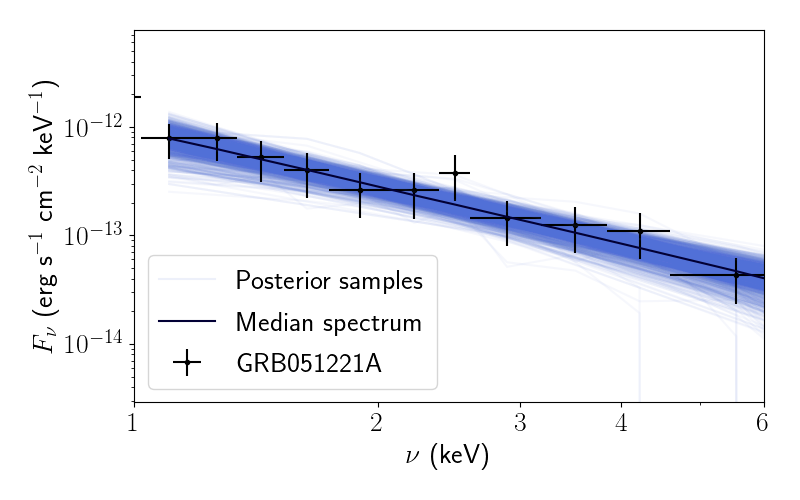}
  \caption{\label{fig:051221A_spectrum} GRB051221A}
  \end{subfigure}
  \begin{subfigure}{0.45\textwidth}
  \includegraphics[width=\linewidth]{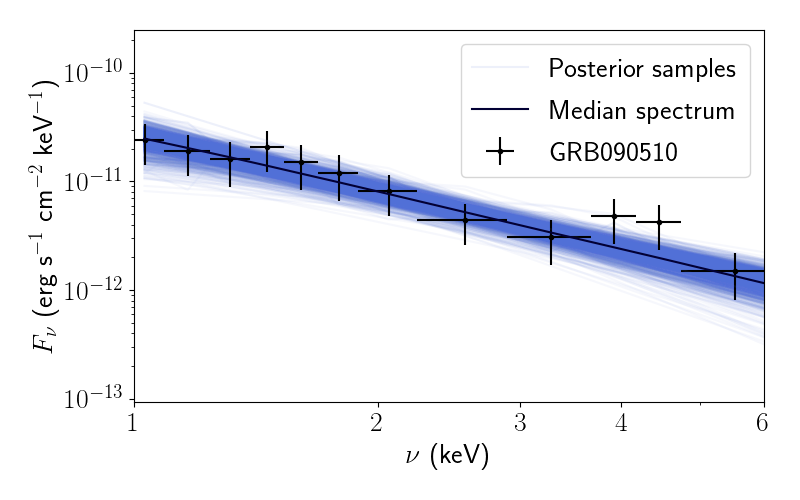}
  \caption{\label{fig:090510_spectrum} GRB090510}
  \end{subfigure}
  \begin{subfigure}{0.45\textwidth}
  \includegraphics[width=\linewidth]{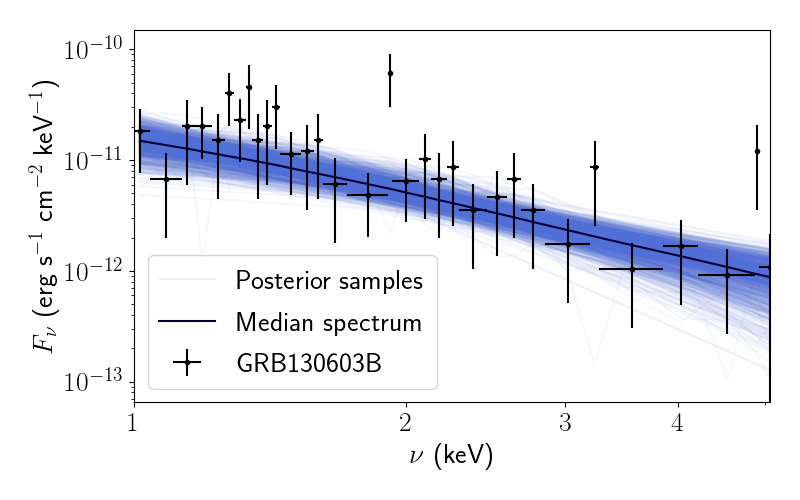}
  \caption{\label{fig:130603B_spectrum} GRB130603B}
  \end{subfigure}
  \begin{subfigure}{0.45\textwidth}
  \includegraphics[width=\linewidth]{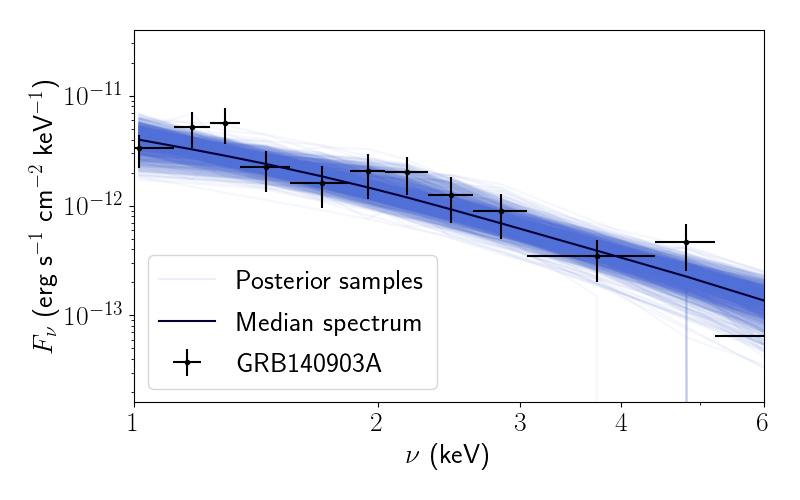}
  \caption{\label{fig:140903A_spectrum} GRB140903A}
  \end{subfigure}
  \begin{subfigure}{0.45\textwidth}
  \includegraphics[width=\linewidth]{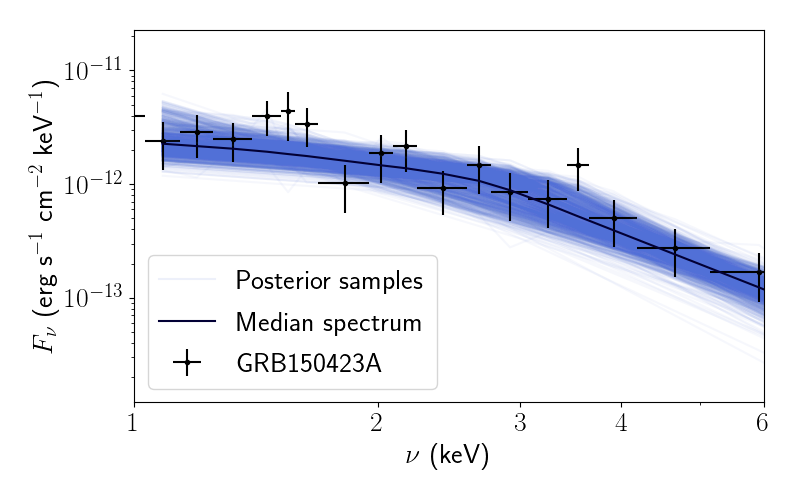}
  \caption{\label{fig:150423A_spectrum} GRB150423A}
  \end{subfigure}
  \begin{subfigure}{0.45\textwidth}
  \includegraphics[width=\linewidth]{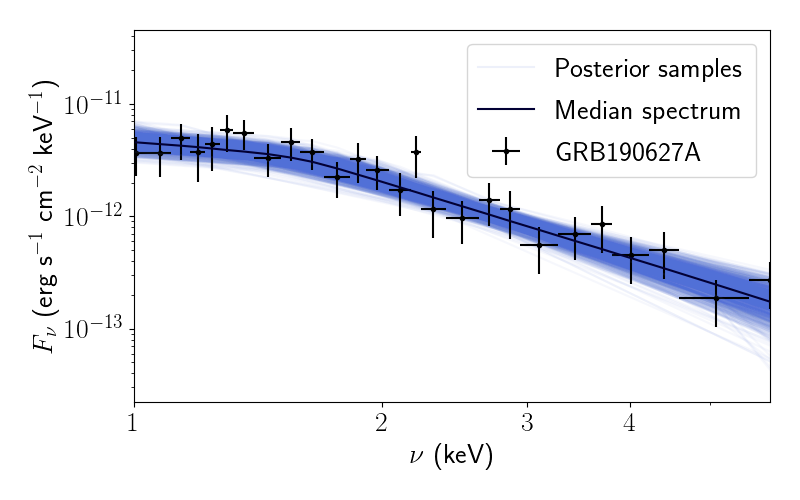}
  \caption{\label{fig:190627A_spectrum} GRB190627A}
  \end{subfigure}
  \caption{\label{fig:spectra} Synchrotron spectral flux density (erg s$^{-1}$ keV$^{-1}$) versus frequency (keV) as a check on the posteriors in Figure \ref{fig:corners} for model A. Black points are data from {\em Swift}; blue curves are spectra produced using 50 random samples from the posterior distributions of each source for model A.}
\end{figure*}

\begin{figure*}
  \begin{subfigure}{0.45\textwidth}
  \includegraphics[width=\linewidth]{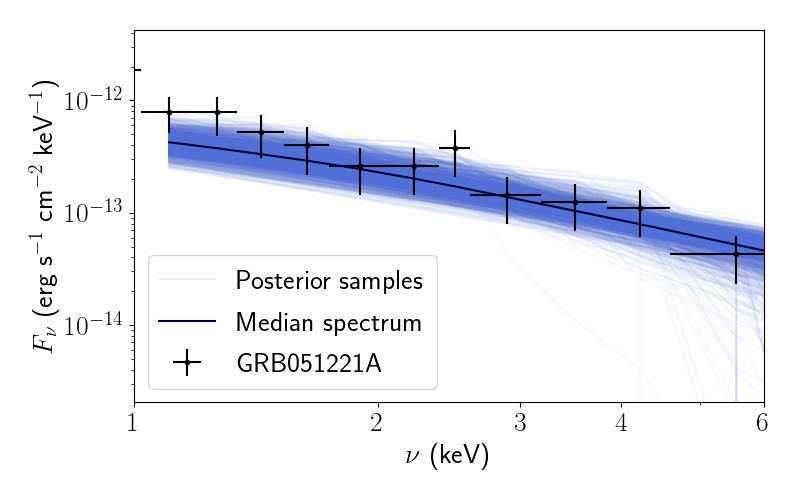}
  \caption{\label{fig:const_051221A_spectrum} GRB051221A}
  \end{subfigure}
  \begin{subfigure}{0.45\textwidth}
  \includegraphics[width=\linewidth]{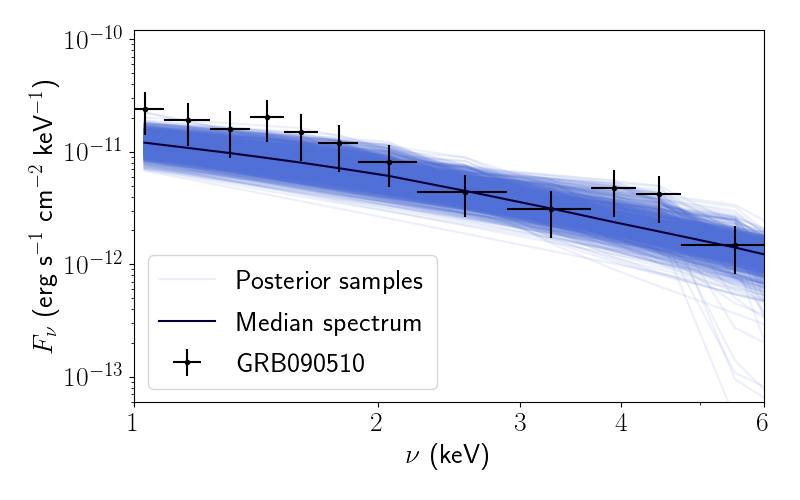}
  \caption{\label{fig:const_090510_spectrum} GRB090510}
  \end{subfigure}
  \begin{subfigure}{0.45\textwidth}
  \includegraphics[width=\linewidth]{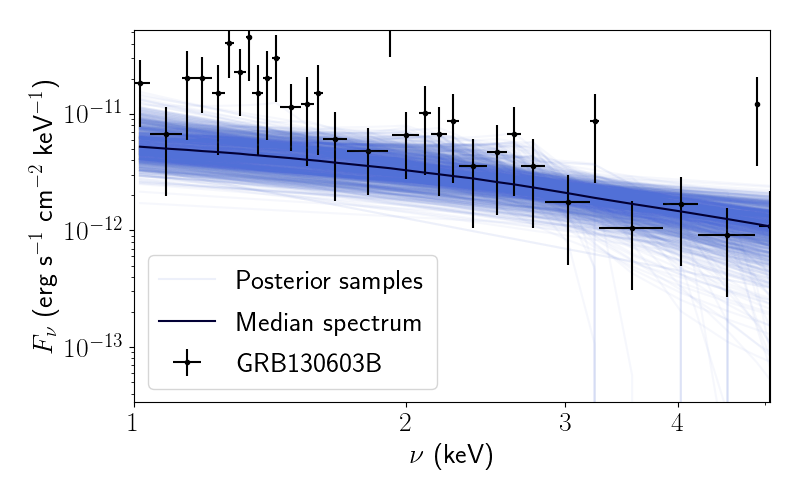}
  \caption{\label{fig:const_130603B_spectrum} GRB130603B}
  \end{subfigure}
  \begin{subfigure}{0.45\textwidth}
  \includegraphics[width=\linewidth]{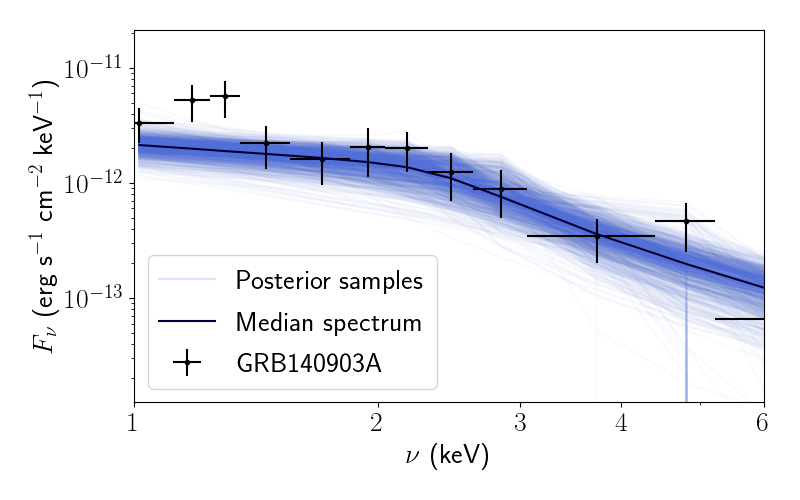}
  \caption{\label{fig:const_140903A_spectrum} GRB140903A}
  \end{subfigure}
  \begin{subfigure}{0.45\textwidth}
  \includegraphics[width=\linewidth]{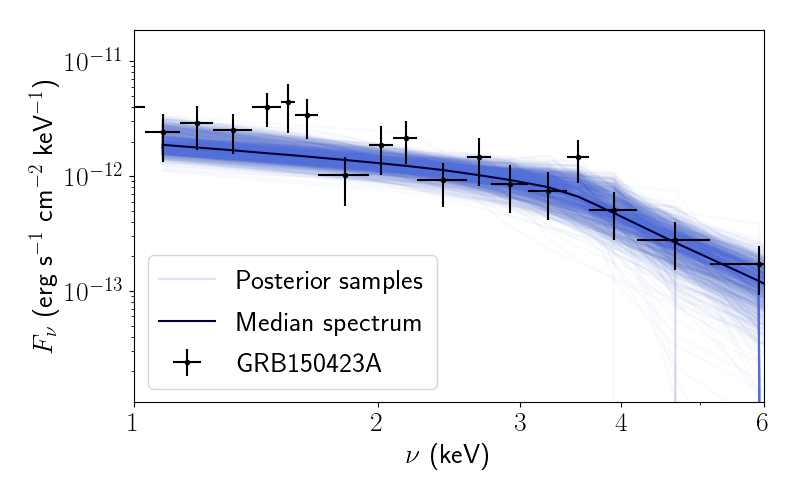}
  \caption{\label{fig:const_150423A_spectrum} GRB150423A}
  \end{subfigure}
  \begin{subfigure}{0.45\textwidth}
  \includegraphics[width=\linewidth]{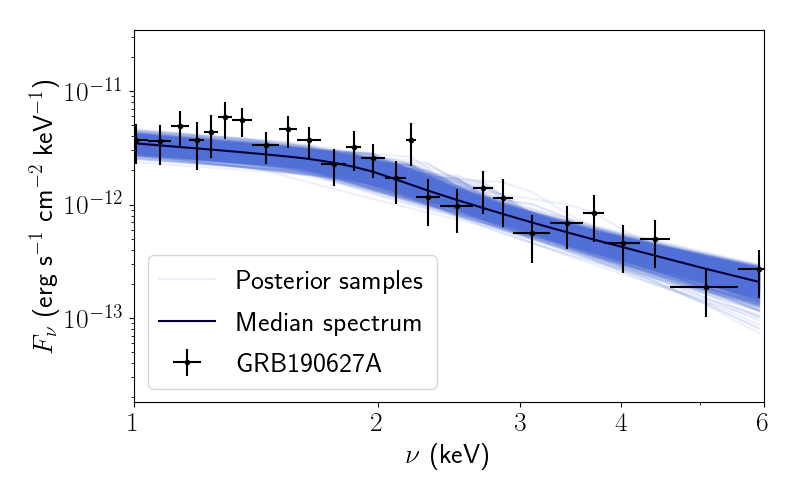}
  \caption{\label{fig:const_190627A_spectrum} GRB190627A}
  \end{subfigure}
  \caption{\label{fig:const_spectra} Same as for Figure 6, but for model B.}
\end{figure*}

We now check that the parameters estimated from the data at $\nu > 1 \, {\rm keV}$ (where absorption is negligible; see appendix \ref{sec:pheldef}) generate spectra which are consistent with the data.
For each sGRB, we randomly sample the posteriors for models A and B produced in Section \ref{sec:results} and calculate theoretical spectra.
We overlay the data with the model predictions for models A (Figure \ref{fig:spectra}) and B (Figure \ref{fig:const_spectra}).
As expected, the theory matches the data for both models.

\section{Spectral evolution}
\label{sec:fullspec}
In this section, we analyse the point-in-time spectrum of GRB130603B at four instants, in the early ($t_1 = 643 \, {\rm s}$ and $t_2 = 745\, {\rm s}$) and late ($t_3 = 5249 \, {\rm s}$ and $t_4 = 5735 \, {\rm s}$) stages of the remnant's evolution, using model B.
The four epochs are defined in Table \ref{tab:fulspec}.
To perform this analysis, we use a joint likelihood, i.e., we define a Gaussian likelihood for each epoch as defined in Table \ref{tab:fulspec} and multiply together the likelihoods for the four epochs, with $\log_{10}B_0$, $\log_{10}\Omega_0/2\pi$, $a$ and $\log_{10}E_{\pm 0}$ being constant at each epoch and $B_i = B(t_i)$.
The constancy of the former four parameters is reasonable physically, as $B_0$ and $\Omega_0$ are set when the central engine forms, and $a$ and $E_{\pm 0}$ are governed by universal aspects of the shock acceleration physics.
In contrast, $\log_{10}B$ is allowed to change from one epoch to the next, with $B_i = B(t_i)$.
This approximation holds, if $N(E,t)$ is quasi-steady in the vicinity of each epoch, in the average sense described in Section \ref{sec:workedex} and checked in Figures \ref{fig:sliced} and \ref{fig:timeave}.
As noted above, the joint likelihood is the product of the Gaussian likelihoods at the four epochs.

\begin{figure*}
  \centering
  \includegraphics[width=\textwidth]{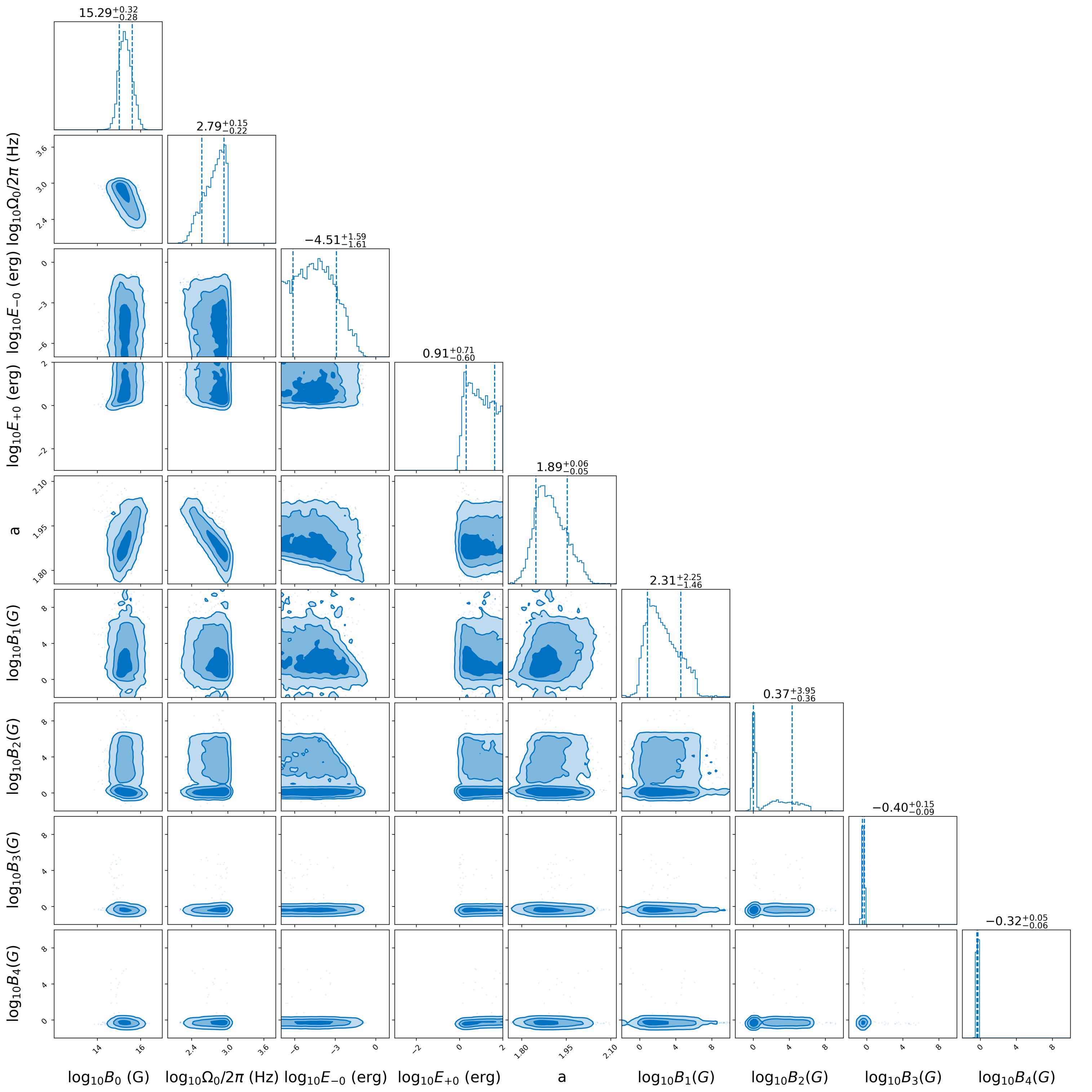}
  \caption{\label{fig:joint} Corner plot showing the posterior distribution obtained for four instantaneous spectra for GRB130603B for the parameters $\log_{10} B_0 \, {\rm (G)}$, $\log_{10}\Omega_0/2\pi \, {\rm (Hz)}$, $\log_{10} E_{-0} \, {\rm (erg)}$, $a$, and $\log_{10} B_i$ ($1 \leq i \leq 4$).}
\end{figure*}

Using the methods described in Section \ref{sec:bayes}, we obtain the posterior displayed as a corner plot in Figure \ref{fig:joint}.
The five parameters that do not evolve ($B_0$, $\Omega_0$, $a$ and $E_{\pm 0}$) return results broadly consistent with the results in Section \ref{sec:results}.
The mean values of the posterior describe a millisecond magnetar with $B_0 \approx 2 \times 10^{15} \, {\rm G}$ and $\Omega_0/2\pi \approx 600 \, {\rm Hz}$, supplying the remnant with a power-law of electrons with power-law index $a \approx 1.9$, $E_{-0} \approx 3.2 \times 10^{-5} \, {\rm erg}$, and $E_{+0} \approx 1 \, {\rm erg}$.
As in Section \ref{sec:params}, the posteriors on the inferred magnetic field $B$ are generally between $10^{-1} \lesssim B/(1 \, {\rm G}) \lesssim 10^3$. 
The median magnetic fields reported in Table \ref{tab:fulspec} suggest the field drops at an average rate of $0.04 \, {\rm G \, s}^{-1}$ from $B_1 = 2 \times 10^2 \, {\rm G}$ at $t_1 = 643 \, {\rm s}$ to $B_4 = 5 \times 10^{-1} \, {\rm G}$ at $t_4 = 5735 \, {\rm s}$.
This is slower than what is expected for the field in the termination shock of the wind in model A, which scales roughly as $B(t)  \propto B_0 t^{-2}$ for $t \gtrsim \tau$, if the wind expands at a constant, relativistic speed \citep{Kennel1984,Strang2019}.
In addition, the hypothetical wind magnetic field $B(t_i) \propto B_0 t_i^{-2}$ in model A is at least one order of magnitude larger than the inferred $B_i$.
This result may point to several possible scenarios.
 \begin{itemize}
    \item The magnetic field in the shock may be dominated by the ambient magnetization of the system instead of the magnetization of the central engine, unlike in Galactic supernova remnants.
The estimates $5\times 10^{-1} \lesssim B_i/({\rm 1 \, G}) \lesssim 2\times 10^2$ are high compared to typical interstellar magnetic fields but low compared to dynamo amplification in the shock or strong fields in the circumstellar environment of the sGRB progenitor.
\item The slow rate of change of the magnetic field in the shock may be explained if the magnetic field is advected outwards by a wind with $B(t) \propto t^{-2}$, if the shock decelerates and stalls behind the merger ejecta.
Dissipation processes may be responsible for reducing $B_i$ below the undissipated split monopole prediction $B(t)  \propto B_0 t^{-2}$ as well.
In this scenario, better time resolution on $B(t)$ would help probe the radial location of the plerion bubble at time $t$.
\item Model B may neglect some critical physics necessary to link the observed synchrotron radiation with the magnetization of the system.
\end{itemize}
At this stage, the data are insufficient to distinguish between the above possibilities (and others).

\begin{table}
\begin{tabular}{lllll}
\hline
Epoch & \(t_{\rm mean}\) (s) & Time span (s) & \(c_f\) (erg cm$^{-2}$ cts$^{-1}$) & $\log_{10}B_i$ (G)\\
\hline
\(t_1\) &  643 &  600-700 & \(6.7\times 10^{-11}\) & $2.3$ \\
\(t_2\) &  745 &  700-800 & \(1.1\times 10^{-10}\) & $0.37$\\
\(t_3\) & 5249 & 5000-5500 & \(5.6\times 10^{-11}\)& $-0.40$\\
\(t_4\) & 5735 & 5500-6000 & \(5.5\times 10^{-11}\) & $-0.32$ \\
\hline
\end{tabular}
\caption{\label{tab:fulspec}
Epochs of four point-in-time spectral snapshots used to jointly analyse the light curve and spectral data from GRB130603B in Section \ref{sec:fullspec}. Here $c_f$ is the counts-to-flux ratio and $t_{\rm mean}$ is the mean photon arrival time, both as reported by the \textit{Swift} online data centre. The last column lists $B_i$ ($1 \leq i \leq 4$) inferred for model B.}
\end{table}

\section{Conclusions}
\label{sec:conclusion}
In this paper we estimate using Bayesian inference the underlying physical parameters of a plerionic model of sGRB X-ray afterglows using data from six sGRBs with known redshifts observed by the \textit{Swift} telescope.
The parameters fall into two categories: those describing the central engine and its magnetized environment ($B_0$, $\Omega_0$, $B$), and those describing the injection of shock-accelerated electrons ($E_{-0}$, $a$).
By analysing point-in-time spectra, we find that the central engine is compatible with a millisecond magnetar, with the posteriors favouring $B_0 \gtrsim 10^{15} \, {\rm G}$ and $\Omega_0/(2\pi) \gtrsim 100 \, {\rm Hz}$.
We also find that $E_{-0}$ and $a$ do not vary over the duration of the X-ray plateau.

We then extend the model to allow the spatially-averaged magnetic field in the synchrotron-emitting bubble to evolve.
Upon analysing the well-sampled spectrum of GRB130603B at four epochs, we infer that $B$ decays slowly from $B_1 = 2 \times 10^2 \, {\rm G}$ at $t_1 = 643 \, {\rm s}$ to $B_4 = 5 \times 10^{-1} \, {\rm G}$ at $t_4 = 5735 \, {\rm s}$.
This result has interesting albeit uncertain implications for the circumstellar environment of the sGRB progenitor and the expansion history of the sGRB blast wave, conditional on the physical ingredients of the plerionic model.

In this paper, we consider only a small sample of sGRBs with X-ray plateaux.
However expanding this sample may be worthwhile.
The analysis in Section \ref{sec:fullspec} should also be extended so as to jointly analyse all the spectral information at all times, by finding a suitable approximation for equation (\ref{eq:spec}) to decrease computational costs.
The analysis prefers bright sources, where one can choose the averaging time-scale $T_{\rm av}$ to be shorter than the natural evolution time-scale of the remnant $\sim 10^2\,{\rm s}$.
A more complete model would also consider radiative transfer through the post-merger shroud for the case $\epsilon \sim 1$.

\section*{Acknowledgements}
Parts of this research were conducted by the Australian Research Council Centre of Excellence for Gravitational Wave Discovery (OzGrav), through Project Number CE170100004.
The work is also supported by Australian Research Council Discovery Project grants (DP170103625) and Discovery Project DP180103155 and Future Fellowship FT160100112.
This work made use of data supplied by the UK Swift Science Data Centre at the University of Leicester.

\section*{Data availability}
This work made use of data supplied by the UK Swift Science Data Centre (\url{https://www.swift.ac.uk/xrt_curves/}) at the University of Leicester.




\bibliographystyle{mnras}
\bibliography{plerion_inference} 



\appendix
\section{Analogy with supernova remnants}
\label{sec:modelcf}

The model used here and presented in \citet{Strang2019} is adapted from the classic model for supernova remnants originating in \citet{Pacini1973}.
In this appendix, we briefly recap the classical supernova remnant model and highlight where the model considered here diverges from it.

The supernova remnant model describes the interaction of a pulsar wind with a thin shell of supernova ejecta at radius $r_{\rm ejecta}$ moving at constant, non-relativistic velocity $v_{\rm ejecta} \ll c$.     
The wind is extremely relativistic and interacts with the ejecta, launching a termination shock at $r_{\rm shock} < r_{\rm ejecta}$.
Here $r_{\rm shock}$ is defined as the radius where the ram pressure $P_{\rm ram}$ in the pulsar wind balances the kinetic pressure $P_{\rm kin}$ in the hot, shocked electrons trapped in the shell $r_{\rm shock} < r < r_{\rm ejecta}$
In this scenario, one finds  $r_{\rm shock}/r_{\rm ejecta} \sim \left(v_{\rm ejecta}/c\right)^{1/2}$ \citep{rees1974origin}. 
Explicitly solving $P_{\rm ram}(r_{\rm shock}) = P_{\rm kin}(r_{\rm shock})$ gives an approximate scaling $r_{\rm shock} \propto t^{-2}$ for $t < \tau$.

\citet{Strang2019} describes the interaction of the pulsar wind with a thin shell formed by a relativistic blastwave with velocity $v_{\rm blast}$.
The pressure in a relativistic shock scales $\propto \left[1-r/(c t)\right]^{-17/12}$ where $r$ is the radial coordinate \citep{Blandford1976}.
This pressure replaces the non-relativistic ejecta in the classical model from \citet{Pacini1973}. 
Again, the interaction launches a termination shock into the pulsar wind which scales as $r_{\rm shock} \propto t^{-2}$ for $t < \tau$.
For an ultrarelativistic shock, this corresponds to $r_{\rm shock}/r_{\rm ejecta} \approx 0.53$.
The region between $r_{\rm shock}$ and $r_{\rm ejecta}$ is a bubble filled with the shocked wind.
The X-ray emission is generated predominantly by freshly-injected, high-energy electrons near $r_{\rm shock}$, which may be approximated as a thin shell.
Further details of the model are presented in \citet{Strang2019}. 

\section{Particle escape}
\label{sec:tesc}

If the neutron star is completely shrouded by a shell of ejecta, the X-ray emission described here is invisible until the ejecta become transparent to X-rays. 
This process has been discussed in detail by previous authors \citep[e.g.][]{Yu2013, Metzger2014,Siegel2016b}.
In this work, we assume that emission from the nebula is able to escape.
This could happen if (for example) the shroud of ejecta is perforated by Rayleigh-Taylor instabilities or pierced by a jet.
In principle, the shocked electrons which produce the observed X-rays could also escape through the holes.
In practice, the electron gyroradius $r_{\rm g} \sim  E/(ceB)$ is less than the thickness of the bubble for all realistic combinations of $E$ and $B$, so the probability of ballistic escape is low. 
However, diffusive leakage through the holes in the shroud is possible.

We present here a simple generalization of the model in Section \ref{sec:model}  that incorporates electron leakage.
The generalization resembles the `leaky box' model used to describe the diffusion of cosmic rays \citep{simpson1983elemental}.
We add a loss term to the right-hand side of  Eq. ~(\ref{eqn:pdegen}), viz.

\begin{equation}
\label{eqn:pdeleak}
    \frac{\partial N(E,t)}{\partial t}  = \frac{\partial }{\partial E} \left[ \left(\left.\frac{dE}{dt}\right|_{\text{ ad }}  + \left.\frac{dE}{dt}\right|_{\text{ syn }} \right) N(E,t)\right] + \dot{N}_{\text{inj}}(E,t) - \frac{N(E,t)}{\tau_{\rm esc}}
\end{equation}
where $\tau_{\rm esc}$ parameterizes the average time taken for an electron to exit the plerion.

This equation can be easily solved for a constant magnetic field (model B) to obtain a Green's function 

\begin{equation}
\label{eqn:greenleak}
G(E, t; t_i) = E^{-2}\left[c_s B^2 (t_i-t) + E^{-1}\right]^{a-2}e^{(t-t_i)/\tau_{\rm esc}}
\end{equation}
where $t_i$ is the injection time of an electron.
Equation (\ref{eqn:greenleak}) differs from the solution in \citet{Strang2019} by the factor $\exp\left[\left(t-t_i\right)/\tau_{\rm esc}\right]$, which reduces to one in the limit $\tau_{\rm esc} \rightarrow \infty$. 
The integral over the injection time $t_i$ can be performed analytically for constant injection ($L_{\rm spin-down}(t) = L_0$).
For $E_{-0} \leq E < E_{+0}$, we find 
\begin{equation}
\begin{aligned}
  N(E,t)  = & \frac{L_0 (2-a)}{E_{+0}^{a-2}-E_{-0}^{a-2}}(c_s B^2)^{a-2} \tau_{\rm esc}^{a-1}E^{-2} \\
            & \times \left\{\Gamma\left[ a-1,-\left(c_s B^2 E \tau_{\rm esc}\right)^{-1} \right ]\right. \\
            & \left. - \Gamma\left[ a-1,-\left(c_s B^2 E_{+0} \tau_{\rm esc}\right)^{-1} \right ]\right\} e^{-\left( c_s B^2 E \tau_{\rm esc} \right)^{-1}}
\label{eqn:leakyn}
\end{aligned}
\end{equation}
where $\Gamma(x,y)$ is the upper incomplete gamma function.
In the limit $\tau_{\rm esc} \rightarrow \infty$, $N(E,t)$ reduces to 
\begin{equation}
N(E,t) = (c_s B^2)^{a-2}E^{-2}L_0 \frac{(a-2)}{a-1}\left(\frac{E^{1-a} - E_{+0}^{1-a}}{E_{-0}^{a-2}-E_{+0}^{a-2}} \right)
\label{eqn:leakynlim}
\end{equation}
A similar result can be obtained for $E < E_{-0}$.
Equation (\ref{eqn:leakynlim}) is identical to the equivalent expression without the leaky box extension applied (see equation (A5) in \citet{Strang2019}). 

\begin{figure}
  \centering
  \begin{subfigure}{0.45\textwidth}
  \includegraphics[width=\linewidth]{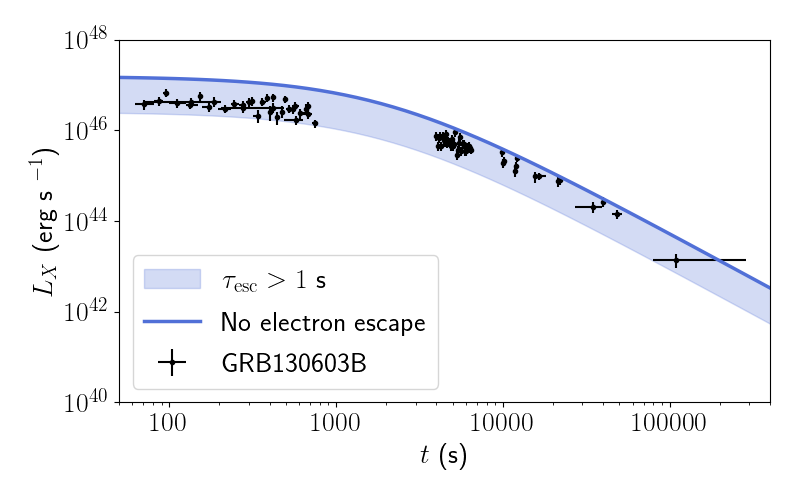}
  \end{subfigure}
  \begin{subfigure}{0.45\textwidth}
  \includegraphics[width=\linewidth]{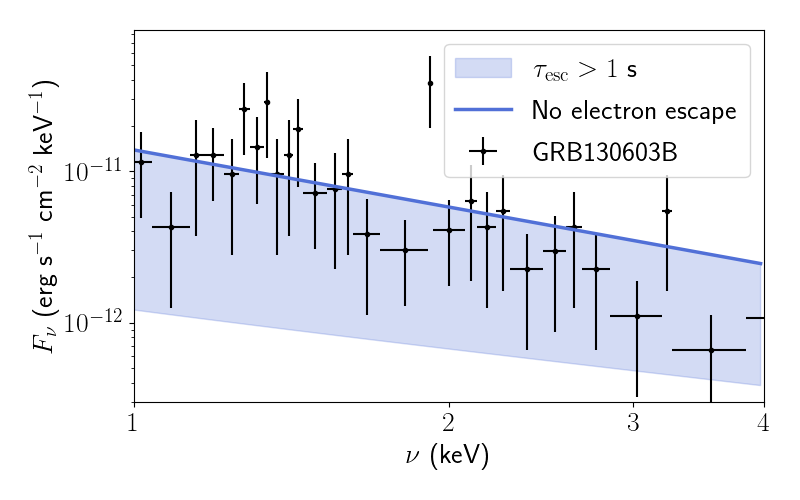}
  \end{subfigure}
  \caption{\label{fig:tesc} Top panel: Synchrotron spectral flux $L_{\rm X}$ (erg s$^{-1}$) versus frequency. Bottom panel: $F_\nu$ (erg s$^{-1}$ cm$^{-2}$ keV$^{-1}$) versus frequency (keV). The light blue region covers the results for $\tau_{\rm esc} > 1 \, {\rm s}$. The dark blue curve is the model with no electron escape. The overplotted black points are the \textit{Swift} observation of GRB130603B. Parameters: $B = 32 \, {\rm G}$, $B_0 =  10^{15} \, {\rm G}$, $\Omega_0/(2\pi) = 775 \, {\rm Hz}$, $E_{-0} = 3\times 10^{-4} \, {\rm erg}$, $E_{+0} = 10 \, {\rm erg}$, and $a = 2.5$}
\end{figure}

We compare the light curves and spectra obtained with and without the leaky box extension.
We calculate both the light curve and spectrum produced by the representative parameter set $B = 32 \, {\rm G}$, $B_0 =  10^{15} \, {\rm G}$, $\Omega_0/(2\pi) = 775 \, {\rm Hz}$, $E_{-0} = 3\times 10^{-4} \, {\rm erg}$, $E_{+0} = 10 \, {\rm erg}$, and $a = 2.5$ for $\tau_{\rm esc} > 1 \, {\rm s}$.
These parameters are consistent with (but not drawn from) the posterior distributions for GRB130603B. 
Figure \ref{fig:tesc} compares the light curve and spectra produced in the range $\tau_{\rm esc} > 1 \, {\rm s}$ (blue shaded region) to the curve with no electron escape (dark blue curve) and overplots the relative observations from \textit{Swift}. 
The overall brightness of the remnant decreases up to $15\%$ with shorter $\tau_{\rm esc}$, however, the variation is within the uncertainty of the observations.
The peak brightness of the spectrum decreases by just $5\%$.
The shape of the light curve and the spectrum are unaffected by $\tau_{\rm esc}$.

\section{Photoelectric absorption}

\label{sec:pheldef}
Frequency-dependent photoelectric absorption from the Milky Way changes the shape of the spectrum, particularly below 1 keV \citep{longair2011high}.
In this appendix, we summarize briefly the effect this has on our inference procedure.
When accounting for absorption, we first calculate the frequency-dependent optical depth $\tau_\nu$ as
\begin{equation}
\tau_\nu = \sigma_k(\nu) n_H,
\label{eqn:opticaldepth}
\end{equation}
where $n_H$ is the hydrogen column density at the sky position of the sGRB.
The photo-electric cross-section $\sigma_K(\nu)$ is

\begin{equation}
  \label{eqn:phxsec}
\sigma_K(\nu) = \sum_i 4\sqrt{2}n_i \sigma_T \alpha^4 Z_i^5 \left( \frac{m_e c^2}{h \nu}\right)^{7/2},
\end{equation}
where $\alpha$ is the fine structure constant and $Z_i$ is the atomic number of the $i^{\rm th}$ element, which has cosmic abundance $n_i$ relative to hydrogen  \citep{Lodders_2003,longair2011high}.
The maximum $Z_i$ is determined by whether enough photons ionize the K-shell  \citep{Burr1967,longair2011high}, which induces sharp jumps in the spectrum as the photon frequency increases.
The precise value of $\sigma_K$ varies by up to a few percent across various analyses \citep{2016ITNS...63.1117H}.
The post-absorption spectrum is then
\begin{equation}
\label{eqn:absorb}
F_\nu^{\rm abs}(t) = e^{-\tau_\nu} F_\nu(t).
\end{equation}

The above approach draws heavily from the model of photoelectric absorption used by the popular X-ray spectral fitting program \texttt{XSPEC} \citep{Arnaud1996}.
In this paper, we neglect the interstellar medium of the host galaxy of the sGRB, due to the paucity of information about its density and composition.
Also, most sGRBs are offset from their host galaxy, so the contribution from the interstellar medium is likely small.
We also neglect absorption in the ``shroud'' of sGRB debris cloaking the central engine.
This shroud is likely to be important, and has been studied by others \citep{Yu2013,Metzger2014,Siegel2016} with respect to absorption and reprocessing of the X-ray emission by the (initially opaque) merger debris.
Following \citet{Strang2019}, we focus on the unshrouded ($\epsilon = 1$) plerionic component, on the grounds that the merger ejecta may not completely conceal the remnant if it is (for example) pierced by a jet or shredded by Rayleigh-Taylor instabilities.
In reality, we expect $\epsilon < 1$.
If $\epsilon$ is constant, it affects the overall normalization of the plerionic component of the remnant luminosity, without affecting the shape of its light curve and spectrum.

\begin{figure}
  \centering
  \includegraphics[width=\columnwidth]{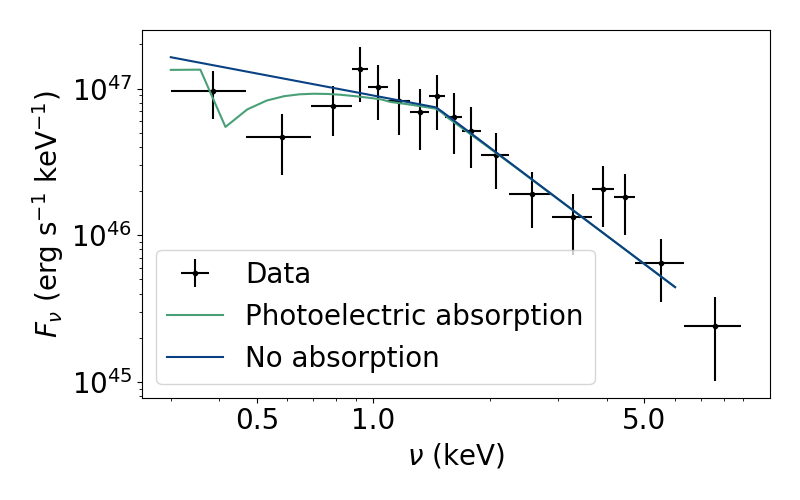}
  \caption{\label{fig:nhcomp} Synchrotron spectral flux density $F_{\nu} \, {\rm (erg \, s}^{-1} {\rm \, keV}^{-1})$ versus observing frequency $\nu \, {\rm (keV)}$. Black crosses are data  from \textit{Swift} for a representative sGRB, GRB090510. The two curves show the plerion model with (green) and without (blue) photoelectric absorption. Parameters: $B_0 = 5\e{15} \, {\rm G}$, $\Omega_0/2\pi = 145 \, {\rm Hz}$, $E_{-0} = 2\e{-4} \, {\rm erg}$, and $a = 3.5$}
\end{figure}

Figure \ref{fig:nhcomp} compares the plerionic emission with and without photoelectric absorption for GRB090510 using parameters guided by (but not drawn from) the results in Section \ref{sec:results}.
The absorption depends strongly on frequency according to equation (\ref{eqn:phxsec}).
At energies $h\nu \lesssim 0.1 \, {\rm keV} $, photoelectric absorption extinguishes the transient.
The band $0.3 \, \lesssim h\nu/(1 \, {\rm keV}) \lesssim 1$ observed by \textit{Swift} contains a break around 1 keV.
The spectrum is flat at $ \nu \lesssim 1 {\rm keV}$ and a power law at $\nu \gtrsim 1$ keV (Figure \ref{fig:nhcomp}).
The sharp drop in the absorbed spectrum at $\approx 0.5 \, {\rm keV}$ is due to the K-edge of oxygen.
The unabsorbed spectrum also shows a break at $\nu \approx 1 \, {\rm keV}$.
The frequency of the latter break depends on the central engine parameters, not photoelectric absorption, and shifts leftwards and upwards with time.

\bsp	
\label{lastpage}
\end{document}